\documentclass{article}
\usepackage[utf8]{inputenc}

\usepackage{amsmath,dsfont,amssymb,array,xcolor,natbib,graphicx,subfigure,hyperref}
\usepackage[left=3cm,right=3cm,top=3cm,bottom=3cm]{geometry}

\newcommand{\N}{\mathbb{N}}

\newcommand{\R}{\mathbb{R}}
\newcommand{\pr}{\mathbb{P}}
\newcommand{\D}{\mathcal{D}}
\newcommand{\B}{\mathcal{B}}
\newcommand{\F}{\mathcal{F}}
\newcommand{\J}{\mathcal{J}}
\newcommand{\Ll}{\mathcal{L}}
\newcommand{\pLl}{\mathcal{\tilde L}}
\newcommand{\Lp}{\mathcal{L}^p}
\newcommand{\p}{\partial}

\newcommand{\X}{\mathbf{X}}
\newcommand{\Nzb}{\eta_0}
\newcommand{\ds}{\displaystyle}

\def\1{\mathds{1}}
\def\lp {\left( }
\def\rp {\right) }
\definecolor{aquamarine}{rgb}{0.13, 0.68, 0.8} 
\newtheorem{lemma}{Lemma}
\newtheorem{corollary}{Corollary}

\newtheorem{remark}{Remark}

\title{Spatial statistics and stochastic partial differential equations: a mechanistic viewpoint}
\author{}
\author{Lionel Roques, Denis Allard, Samuel Soubeyrand \\
\small{Biostatistics and Spatial Processes (BioSP), INRAE, 84914 Avignon, France}}

\date{}

\begin{document}

\maketitle

\begin{center}
\begin{minipage}{0.8\linewidth}

\noindent {\textbf{Abstract}}

The Stochastic Partial Differential Equation (SPDE) approach, now commonly used in spatial statistics to construct Gaussian random fields, is revisited from a mechanistic perspective based on the movement of microscopic particles, thereby relating pseudo-differential operators to dispersal kernels. We first establish a connection  between L\'evy flights and PDEs involving the Fractional Laplacian (FL) operator. The corresponding Fokker-Planck PDEs will serve as a basis to propose new generalisations by considering a general form of SPDE with terms accounting for dispersal, drift and reaction.  We detail the difference between the FL operator (with or without linear reaction term) associated with a fat-tailed dispersal kernel and therefore describing long-distance dependencies, and the damped FL operator associated with a thin-tailed kernel, thus corresponding to short-distance dependencies. Then, SPDE-based random fields with non-stationary external spatially and temporally varying force are illustrated and nonlinear bistable reaction term are introduced.  The physical meaning of the latter and possible applications are discussed. Returning to the particulate interpretation of the above-mentioned equations, we describe in a relatively simple case their links with point processes. We unravel the nature of the point processes they generate and show how such mechanistic models, associated to a probabilistic observation model, can be used in a hierarchical setting to estimate the parameters of the particle dynamics.

\medskip

\end{minipage}
\end{center}

\section{Introduction}

Diffusion processes are commonly used for modelling natural space-time processes in climate sciences, ecology, epidemiology and population dynamics, to name a few. For example, the diffusion equation, $\partial_t u(x,t)  = \Delta u(x,t)$ with $x \in \mathbb{R}^2, t > 0$, one of the most elementary partial differential equations (PDEs) known as the heat equation, describes the macroscopic behaviour of many micro-particles resulting from their random movements. How PDEs such as the heat equation, but also more complex ones, interact with spatial statistics is the focus of this contribution.

PDEs and statistics have long been connected through the usual question of parameter estimation in mechanistic models \citep[see e.g.,][and the references therein]{Wik03b,SouRoq14}. About ten years ago, \citet{LinRue11} revisited the explicit link between stochastic  partial differential equations (SPDEs) and spatial statistics already established in \citet{Whi54,Whi63}. They introduced a new paradigm for the statistical analysis of spatial data that consists in modelling a spatial variable as the solution of a certain class of SPDE for which the representation on a finite grid presents interesting Markov properties, making it computable for very large grids and capable of handling very large datasets. This approach was based on the observation made in \citet{Whi54,Whi63} that, on $\mathbb{R}^d$, the solution of the SPDE
\begin{equation}
	(\kappa^2 - \Delta)^{\alpha/2}U = W_S, 
	\label{eq:Lindgren2011}
\end{equation}
is a Gaussian random field whose covariance function is proven to be a Mat\'ern covariance. In \eqref{eq:Lindgren2011} $\alpha > d/2$, $\Delta$ is the usual Laplacian, $W_S$ is a Gaussian (spatial) White Noise process and $U$ is the unknown stationary random field. For $\alpha \neq 2$, the pseudo-differential operator in \eqref{eq:Lindgren2011} was constructed on an ad hoc basis to get a Mat\'ern covariance. As \cite{Whi54} himself indicated ``it is difficult to visualize a physical mechanism which would lead to \eqref{eq:Lindgren2011}". 
However, in contrast to the previous statistically oriented constructions, the SPDE approach should also lead to a physically grounded construction, for which the  parameters would carry traditional physical interpretation such as diffusivity, reaction and transport. The SPDE approach also allows the construction of models with interesting properties, such as non-symmetry and space-time non-separability  \citep{brown2000blur}.  Non-stationarity can also easily be accounted for by letting the parameters in \eqref{eq:Lindgren2011} be space-dependent, see e.g. \citet{LinRue11} and \citet{fuglstad2015exploring}. Space-time generalisation is not straightforward. In particular, the space-time covariance is not always accessible (but the spectral density is) and the definition of the space-time white noise requires some care. Recent advances have been made in \citet{bakka2020diffusionbased} and  \citet{CarAll21}, where space-time models admitting Matérn covariance functions in space have been proposed.

\cite{CarAll21} proposed a rigorous mathematical framework, based on the theory of Generalized Random Fields \citep{ito1954stationary, gelfand1964generalized}, to construct spatio-temporal random fields arising from a very broad class of linear SPDEs. They considered the general class of linear SPDEs of the form $\Ll_g U = X$ defined through operators of the form
\begin{equation}
\Ll_g(\cdot):= \F^{-1}(g \F(\cdot)),
\label{eq:Lg}
\end{equation}
where, in the real case, $g:\R^d \to \R$ is a continuous, symmetric function bounded by a polynomial,  $\F$ the Fourier transform on $\R^d$  and $\F^{-1}$ the inverse Fourier transform.  The function $g$ in \eqref{eq:Lg} is referred to as the {\em symbol function} of the operator.  If, in addition, $|g|$ is bounded from below by the inverse of a strictly positive polynomial, there exists a unique Generalized Random Field solution to the equation $\Ll_g U \stackrel{\mathrm{2nd. o.}}{=} X,$  where `2nd. o.' means that the solution has to be understood in the second order sense \citep[see][for more details]{CarAll21}.  In this case, the spectral measures of $X$ and $U$ are connected by:$$d\mu_X=|g|^{2}d\mu_U^X,$$which leads to the covariance $\rho_U=\F(|g|^{-2}\, d\mu_X)$. Thus,  the  relationship between the covariances of $X$ and $U$ is 
$$\rho_{X} = \Ll_{|g|^2} \rho_U.$$
When $X$ reduces to a white noise $W$, its spectral measure is  $d\mu_X(\xi)=(2\pi)^{-d/2} \, d\xi$, and the covariance $\rho_X$ reduces to the Dirac $\delta$. Hence, the covariance $\rho_U$ can be seen as a Green's Function of the operator $\Ll_{|g|^2}$. This construction offers the possibility to build and characterize models far beyond the Matérn family which is currently the covariance model considered within most SPDE implementations.

In this contribution, our aim is to further explore the intimate interplay between partial differential equations, their stochastic extensions, namely the SPDEs, and spatial statistics. Most of the literature about PDEs focused on elliptic and parabolic operators involving a Laplacian or its anisotropic extension. Although the fractional Laplacian has recently generated much attention in the PDE community, its microscopic interpretation is much less classical than that of the standard Laplacian. As a preliminary to this study, we propose in Section~\ref{sec:Levy} an introduction to the connection between L\'evy flights and PDEs involving the fractional Laplacian operator. 
The corresponding Fokker-Planck PDEs will serve as a basis to revisit in Section~\ref{sec:SPDES} the original SPDE framework associated with Eq.~\eqref{eq:Lindgren2011},  but with a mechanistic perspective.  We propose new generalisations by considering a general form of SPDE with terms accounting for dispersal, drift and reaction. Regarding the dispersal term, we first show that there is an equivalent integral representation offering a mechanistic interpretation of the operator $-L^{\alpha/2}:=-(\kappa^2-\Delta)^{\alpha/2}$  in the left-hand-side of Eq.~\eqref{eq:Lindgren2011}. As a side result, we show the profound difference between the fractional Laplacian $-(-\Delta)^{\alpha/2}$, which is associated with a fat-tailed dispersal kernel (for $\alpha\in (0,2)$) and therefore describes long-distance dependencies, and the damped fractional Laplacian $-L^{\alpha/2}$, which is associated with a thin-tailed kernel, corresponding to short-distance dependencies.  We then introduce the fractional Laplacian operator with a linear reaction term $-(-\Delta)^{\alpha/2}-\kappa^2$ and we also consider kernel-based dispersal operators. In each case, we derive  the associated  covariance function and we compare its properties with those of the Mat\'ern family. We then illustrate SPDEs with non-stationary drift terms representing external spatially and temporally varying forces.  Finally, we introduce nonlinear bistable reaction terms~; we discuss their physical meaning in a deterministic framework and the possible applications of the corresponding SPDE, the stochastic Allen-Cahn equation.
Lastly, in Section~\ref{sec:PDEs}, we revisit the classical use of statistical models for parameter estimation in PDE models, with which we started this introduction. In  this original approach, we combine the Fokker-Planck PDEs of Section~\ref{sec:Levy} with inhomogeneous point processes. Section~\ref{sec:disc} concludes with some elements of discussion, with a particular focus on the form of the diffusion operator.

In the following,  $U(x,t)$ will represent a random field defined over  space and time, with  $x = (x_1,\dots,x_d) \in \mathbb{R}^d$ and $t \in \mathbb{R}^+$.   Partial derivatives along time will be denoted $\partial_t$ and $\Delta = \sum_{i=1}^d \partial^2 / \partial x_i^2$ is the Laplacian. The Fourier transform is defined, as follows: ${\cal F}(f)(\xi) = (2\pi)^{-d/2} \int_{\mathbb{R}^d} e^{-i\xi \cdot  x} f(x) dx$.  As already mentioned in the first paragraph, we will often adopt a microscopic point of view, in which solutions of PDEs or SPDEs describe the density of  particles subject to diffusion equations.

\section{L\'evy flights and PDEs with fractional Laplacian \label{sec:Levy}}
Since Einstein's theory of Brownian motion \citep{Ein05}, the parabolic PDE $\p_t u (x,t)= D \Delta u(x,t)$ has a well-established physical interpretation. The solution $u(x,t)$ can indeed be seen as the density of particles following independent Brownian motions. More generally, 
Fokker-Planck equations act as a natural bridge between parabolic PDEs and It\^o diffusion stochastic differential equations (SDEs) \citep[e.g.,][]{Gar09}. These SDEs extend Brownian motion by introducing a drift (or similarly a transport) coefficient and spatio-temporal heterogeneities in the movement of particles. The solution of parabolic-like equations but with a fractional Laplacian $-(-\Delta)^{\alpha/2}$ (with $\alpha\in (0,2)$) instead of a standard Laplacian ($\alpha=2$) can also be interpreted as a density of particles. But in this case the particles follow independent L\'evy flights. As this physical interpretation is less standard, we propose below a short introduction to the connection between L\'evy flights and Fokker-Planck PDEs involving a fractional Laplacian.

We consider a particle whose position $X_t\in \R^d$ follows a  $d-$dimensional $\alpha-$stable L\'evy process, for some $\alpha\in (0,2]$ \citep[see e.g.,][for precise definitions]{ConTan03,App09}. This can be described by a Langevin-like equation \citep[e.g.][]{SchLar01,App09}:
\begin{equation} \label{eq:levydiffusion}
d X_t= B(t,X_t) \,  dt + \Sigma(t,X_t) dL_t,
\end{equation}
where $L$ is an $\alpha-$stable rotation-invariant $d$-dimensional L\'evy process with scale parameter $\gamma$ and $dL_t$ is a forward increment in time of this process. As mentioned above, when $\alpha=2$, $L$ is a standard Brownian motion and the equation \eqref{eq:levydiffusion} is a standard It\^o diffusion SDE  \citep[for a discussion of the differences between It\^o and Stratonovitch SDEs, the reader can refer to][]{Gar09,HorLef06}. The vector field $B(t,X_t)=(B_1(t,X_t), \ldots, B_d(t,X_t))\in \R^d$ is the drift coefficient. It is related to the existence of some external force acting on the particles, sometimes also called {\em bias}, e.g. due to wind or attraction/repulsion. For simplicity, we assume an isotropic  and time-independent random driving force (but see Remark~\ref{rem:aniso} below), which means that the term $\Sigma$ (the diffusion matrix when $\alpha=2$) satisfies $\Sigma(t,X_t)=\sigma(X_t)\, I_d$ where $\sigma$ is a scalar function and $I_d$ is the identity matrix of size $d$. Note that, in general, the coefficients $B$ and $\Sigma$ may depend on the current position $X_t$. For example, the particle motion could be slower in some regions of space, which would be modelled with a smaller value of $\sigma$ in these regions. Note that the existence and uniqueness of the solution of \eqref{eq:levydiffusion} are guaranteed under some technical growth conditions on the coefficients \citep[see e.g.,][]{Arn74,SchLar01,Oks03}.

The Fokker-Planck equation describes the dynamics of the transition probability density $P(X(t)=x|X(0)=x_0)$. We denote by $p(x,t)$ this transition probability. In the standard diffusion case $\alpha=2$,
the dynamics of $p(x,t)$ are described by the parabolic PDE \citep[see e.g.][]{Gar09}:
\begin{align}\label{eq:FP_diff}
\partial_t p (x,t)=  \frac{1}{2}\Delta (\sigma^2(x) \,  p (x,t)) - \hbox{div}\lp B(x,t)\, p (x,t)\rp, \ t>0, \ x\in\R^d, 
\end{align}
where the divergence operator `$\hbox{div}$' acts on the vector fields $V\in C^1(\R^d,\R^d)$, $\hbox{div}(V):=\sum_{i=1}^d \partial V_i / \partial x_i.$ When the movement of the particles is not driven by a Gaussian noise (i.e. when $\alpha<2$), the Fokker-Planck equation involves a fractional Laplacian \citep{SchLar01}:
\begin{equation}\label{eq:FP_Levy}
\partial_t p (x,t)= -\gamma\, (-\Delta)^{\alpha/2}(\sigma^\alpha(x) \,  p (x,t)) - \hbox{div}\lp B(x,t)\, p (x,t)\rp , \ t>0, \ x\in\R^d,
\end{equation}
see Section~\ref{sec:SPDES} for more details about this operator; see also \cite{Kwa15} for alternative definitions of this operator.  Notice that, with $\alpha=2$ and $\gamma=1/2$ (Brownian diffusion), the equation \eqref{eq:FP_Levy} is equivalent to \eqref{eq:FP_diff}.

\begin{remark}\label{rem:aniso}
In the general anisotropic case, when $\alpha=2$ and $\Sigma(t,X_t)$ is not necessarily of the form $\sigma(X_t)\, I_d$, we obtain an equation similar to \eqref{eq:FP_diff}, but the term $ \frac{1}{2}\Delta (\sigma^2(x) \,  p (x,t))$ must be replaced by $\sum_{i,j=1}^d \p^2/\p_{x_i}\p_{x_j}(D_{ij}(x,t) p(x,t))$ with $D_{ij}$ the entries of the matrix $\mathbf{D}(x,t):=\Sigma \, \Sigma^{\tiny\hbox{T}}/2$. When $\mathbf{D}$ does not depend on $x$, we note that this operator reduces to $\hbox{\normalfont div}( \mathbf{D}(t) \nabla p(x,t))$. Otherwise, it can be written as a sum of a `Fickian' diffusion term (see also the discussion in Section \ref{sec:disc}) and a drift term: $\sum_{i,j=1}^N \p^2/\p_{x_i}\p_{x_j}(D_{ij}(x,t) p(x,t))=\text{\normalfont div}( \mathbf{D}(x,t) \nabla p(x,t))+\text{\normalfont div}[\text{\normalfont div}(\mathbf{D}(x,t)) p(x,t)]$, with $\text{\normalfont div}(\mathbf{D}(x,t))$ the vector with coordinates $\sum_{i=1}^N \p/\p_{x_i}D_{ij}(x,t)$.
\end{remark}

The transition probability density $P(X(t)=x|X(0)=x_0)$ involves an initial condition $\delta(x-x_0)$, where $\delta$ is a Dirac measure at $0$. Assuming $N_0$ independent particles, with common initial distribution $p_0$, it is then straightforward to check that the expected density of  particles $u(x,t)$ satisfies the PDE \eqref{eq:FP_Levy}, with the initial condition $u_0(x):=u(x,0)=N_0\, p(x,0).$ 

Assume further that the particles have a life expectancy $1/\kappa^2$ for some $\kappa>0$ and disappear at an exponential rate. Consider any given particle. We denote by $\tau$ its ``death" time (or removal time: we not only focus on living organisms). With a slight abuse of language, we define $h(x,t)$ as the probability that the particle is at the position $x$ and still dispersing at time $t$, that is, for any measurable set $\Omega \subset \R^d$,
$$\int_\Omega h(x,t) dx:=P(\{X_t\in \Omega \}\cap \{t <\tau\}).$$Since dispersal and death are defined as two independent processes, $
h(x,t)= p(x,t) \, P(t<\tau)=p(x,t)\, e^{- \kappa^2 \, t}$. Therefore, $h(x,t)$ and consequently the expected particle density $u(x,t)$ satisfy the following equation:
\begin{equation}\label{eq:FP_Levy_u}
\partial_t u (x,t)= -\gamma\, (-\Delta)^{\alpha/2}(\sigma^\alpha(x) \,  u (x,t)) - \hbox{div}\lp B(x,t)\, u (x,t)\rp- \kappa^2\, u(x,t), \ t>0, \ x\in\R^d.
\end{equation}

This construction highlights two interesting connections between PDEs and stochastic processes. On the one hand, it illustrates that some PDEs with fractional Laplacian arise quite naturally as the Fokker-Planck equation of SDEs with Lévy flights. On the other hand, the PDE in \eqref{eq:FP_Levy_u} suggests interesting generalisations of the SPDE in \eqref{eq:Lindgren2011} which will be the subject of the next section.

\section{SPDEs and random fields: a mechanistic point of view \label{sec:SPDES}}

Inspired by \eqref{eq:FP_Levy_u}, we consider a quite general class of time-dependent SPDE models that rely on mechanistic assumptions concerning the underlying variable. These models take the following general form:
\begin{equation}\label{eq:SPDE_gale}
    \p_t U(x,t)= \underbrace{\D(x,t,[U])}_{\text{dispersal}} +\underbrace{\B (x,t,[U])}_{\text{drift}}+ \underbrace{f(x,t,U)}_{\text{reaction}} + \underbrace{\sigma\,  W_t(x)}_{\text{noise}}, \quad (x,t) \in \R^d \times \R^+.
\end{equation}
The random field $U(x,t)$ is thus governed by a dispersal term $\D$, a drift term $\B$, a reaction term $f$ describing the local feedback of the random field $U$ on itself and a spatio-temporal white noise $\sigma\,  W_t$ defined as the derivative of a Wiener process, and thus verifying $E[W_t(x)W_{t'}(x')]=\delta(t-t')\delta(x-x')$.  The square bracket symbol $[U]$ is used to underline that the operator may depend on the whole function $U(\cdot,t)$, and not only on its value at $x$. The well-posedness of \eqref{eq:SPDE_gale} depends on its precise form and on the space dimension $d$. When the reaction term is linear and with coefficients that are constant in $(x,t)$, the existence and uniqueness of a stationary distribution-valued solution generally follow  from the results in \citet{CarAll21}, at least in the examples that we treat below.
With nonlinear reaction terms $f(x,t,U)$ (nonlinear with respect to $U$), the theory is also well-established in the case $d=1$ \citep[e.g.][]{DapZab14}. The picture is less clear   when $f(x,t,U)$ is nonlinear and $d\ge 2$: although such equations are regularly used in applied sciences, mathematical results tend to show that additive white noise leads to ill-posed equations \citep{HaiRys12,RysNig12}.

In the particular case  $\D(x,t,[U])=D\, \Delta U(x,t)$,   $\B(x,t,[U])=0$ (no drift) and  $f(x,t,u)=-\kappa^2 \, U(x,t)$ for some $D,\kappa>0$ the solution $U(x,t)$ of \eqref{eq:SPDE_gale} satisfies the ``standard" evolving Mat\'ern equation, already discussed in \citet{LinRue11} and \citet{CarAll21}:
\begin{equation}\label{eq:basic}
    \p_t U(x,t)= \underbrace{D\Delta U(x,t)}_{\text{dispersal}}  \underbrace{-\kappa^2 \, U (x,t)}_{\text{reaction}} + \underbrace{\sigma\,  W_t(x)}_{\text{noise}}, \quad (x,t) \in \R^d \times \R^+, \tag{$\mathcal{S}$}
\end{equation}
In this equation, there is a diffusive dispersal term and an absorption term (the reaction) which forces the solution to remain close to $0$, see Section~\ref{sec:Levy} for a microscopic interpretation. When the noise does not depend on time (spatial white noise $W_S$), 
the SPDE \eqref{eq:Lindgren2011} in \cite{LinRue11} is  a time-independent solution of \eqref{eq:basic} for $\alpha=2$. In the general case $\alpha\neq 2$ the SPDE \eqref{eq:Lindgren2011} involves the ``damped" fractional  operator $-L^{\alpha/2}:=-(\kappa^2-\Delta)^{\alpha/2}$, which does not exactly fit in \eqref{eq:SPDE_gale}. 

We will discuss below three extensions of the ``standard" SPDE-GMRF that go beyond the framework \eqref{eq:basic} set in \citet{LinRue11} and their possible applications in environmental and ecological sciences. These extensions either deal with the dispersal term (Section \ref{sec:dispersal}), the drift term (Section \ref{sec:drift}) or the reaction term (Section \ref{sec:bistable}).

\subsection{Dispersal operators} 
\label{sec:dispersal}

In equations of the form \eqref{eq:SPDE_gale}, nonlocal dispersal operators
\begin{equation}\label{eq:D_kernel}
\D(x,t,[U])=D\, (\J\star U- U)(x,t):=D\int_{\R^d}\J(x-y)(U(y,t)-U(x,t)) \, dy,
\end{equation}
can describe short or long-distance dispersal, depending on the precise shape of the kernel $\J$ \citep{KleLav06,Gar11}.  When imposing that the integral of $\J$ is equal to 1, the quantity $\J(x-y)$ can be interpreted as the probability density that a dispersing particle initially located at $y$ moves to the position $x$ \citep[see e.g.,][]{Roq13}, and the coefficient $D>0$ is the rate of dispersal. The frequency of long-distance dispersal events is governed by the tail of the kernel  as $\|x\|\to \infty$:
thin-tailed kernels decay at least exponentially and fat-tailed kernels decay more slowly than any exponential function.

Equations of the form \eqref{eq:SPDE_gale} with dispersal terms of the form \eqref{eq:D_kernel} and regular kernels $\J$ of mass $1$, null drift term  $\B(x,t,[U])=0$  and  $f(x,t,U)=-\kappa^2 \, U(x,t)$ fit in the class of SPDEs considered in \citet{CarAll21}, as $-D\, (\J\star U- U)+\kappa^2 \, U= \F^{-1}(g \, \F(U))$, with symbol function $g=D(1-(2\pi)^{d/2}\F(\J))+\kappa^2$, which is an Hermitian-symmetric function \citep[see Theorem~1 in][]{CarAll21}. We also consider  singular kernels for which, even if $\J\star u$ is not properly defined, the right-hand side in  \eqref{eq:D_kernel} is still well-defined, for instance when $\D(x,t,[U])=-(-\Delta)^{\alpha/2}U(x,t)$. We compare here, from a mechanistic viewpoint, the physical interpretation of these equations
\begin{equation}\label{eq:convo}
    \p_t U(x,t)= D\, (\J\star U- U)(x,t)  -\kappa^2 \, U (x,t) + \sigma\,  W_t(x),
\end{equation}
compared to evolving Mat\'ern equations 
\begin{equation}
    \label{eq:evol_Matern}
    \p_t U(x,t)= -L^{\alpha/2}U(x,t) +\sigma\,  W_t.
\end{equation}
 The main findings of this section are summarized in Table~\ref{tab:ccl}.

\subsubsection{Fractional Laplacian}
We begin by recalling that the fractional Laplacian $-(-\Delta)^{\alpha/2}$ has a well-known pointwise representation when $\alpha\in (0,2)$ \citep[e.g.,][]{Kwa15}: for all $v \in \Lp (\R^d)$ ($p\in [1,\infty)$),
\begin{equation} \label{eq:pointwise_frac_laplace}
    -(-\Delta)^{\alpha/2}v(x)=c_{d,\alpha}\int_{\R^d}\frac{v(y)-v(x)}{\|x-y\|^{d+\alpha}} \,d y,
\end{equation}
for some constant $c_{d,\alpha}>0$. Due to a singularity around $0$, the above integral has to be understood in the Cauchy principal value sense when $\alpha\in [1,2)$. Equation \eqref{eq:pointwise_frac_laplace} shows that the 
fractional Laplacian is a nonlocal operator, and formally corresponds to a dispersal term of the form \eqref{eq:D_kernel}, although the kernel 
$\J(x)=c_{d,\alpha}/\|x\|^{d+\alpha}$ is not integrable around $0$. The kernel $\J$ belongs here to the family of (very) fat-tailed kernels, which is consistent with the microscopic interpretation of this operator as seen in Section~\ref{sec:Levy}, since L\'evy flights \eqref{eq:levydiffusion} are continuous time random walks, with possible large jumps if $\alpha<2$.

\subsubsection{Fractional damped Laplacian} 

We recall that the solution to $-L^{\alpha/2}U(x) = W_S$ in dimension $d \ge 1$ is a Gaussian random field with  Mat\'ern covariance of parameter $\nu=\alpha-d/2  > 0 $:
\begin{equation}
    \label{eq:matern_def}
    \rho_U^{W_S}(h)=\frac{1}{(2\pi)^{\frac{d}{2}}2^{\alpha-1}\kappa^{2\alpha-d}\Gamma(\alpha)}(\kappa\, \|h\|)^{\alpha-\frac{d}{2}}K_{\alpha-\frac{d}{2}}(\kappa \|h\|), \ h\in \R^d,
\end{equation}
where $K_\nu(\cdot)$ is the modified Bessel function of the second kind with index $\nu > 0$.  Notice that in the limit case $\nu=0$, $\rho_U^{W_S}(h)$ is a generalized Matérn covariance  which is well defined for $\|h\| >0$.

In order to find kernel ${\cal J}$ associated to $-L^{\alpha/2}$,   we use a semigroup formula  \citep[e.g.,][]{Sti19} (for $\alpha \in (0,2)$ and $v\in \Lp (\R^d)$, $p\in [1,\infty)$):
\begin{equation}\label{eq:semigroup}
    -L^{\alpha/2}v=\frac{1}{|\Gamma(-\alpha/2)|}\int_0^\infty \lp e^{-t \, L} v - v \rp \frac{1}{t^{1+\alpha/2}} \, dt,
\end{equation}
with $\Gamma$ the gamma function. The quantity $u(\cdot,x)=e^{-t \, L} v $ 
is the semigroup generated by $L$ acting on $v$, i.e., the solution of the Cauchy problem:
$$\p_t u=-L \, u=\Delta \, u - \kappa^2 \, u, \quad t>0, \ x\in \R^d, \hbox{ with }u(\cdot,0)=v.$$
More explicitly,  $u(x,t)=e^{-\kappa^2 \, t }(G_t \star v)(x)$, with $G_t(x)=(4\pi t)^{-n/2}\, \exp(-\|x\|^2/(4\, t)),$ the heat kernel. Replacing into \eqref{eq:semigroup}, we have:
$$
    -L^{\alpha/2}v(x)=\frac{1}{|\Gamma(-\alpha/2)|}\int_0^\infty \lp \int_{\R^d}  e^{-\kappa^2 \, t }G_t(x-y)v (y)\, dy -v(x) \rp\frac{1}{t^{1+\alpha/2}} \, dt.
$$
Thus, 
\begin{align} \label{eq:dbint}
    -L^{\alpha/2}v(x)& =\frac{1}{|\Gamma(-\alpha/2)|}\int_0^\infty \int_{\R^d}  e^{-\kappa^2 \, t }G_t(x-y)(v (y)-v(x)) \frac{1}{t^{1+\alpha/2}} \, dy \, dt - v(x) \, h_\kappa,
\end{align}
with 
\begin{equation} \label{eq:hkappa}
     h_\kappa:= \frac{1}{|\Gamma(-\alpha/2)|}\int_0^\infty  \lp  1-e^{-\kappa^2 \, t }  \rp  \frac{1}{t^{1+\alpha/2}}\, dt \in (0,\infty) \hbox{ for }\alpha \in (0,2).
\end{equation}
Then, we note that:
\begin{align*}
  \int_0^\infty e^{-\kappa^2 \, t }G_t(x-y) \frac{1}{t^{1+\alpha/2}}  \, dt 
    & = \frac{2\, (2\kappa)^{(\alpha+d)/2}}{(4\pi)^{d/2}} \frac{K_{(\alpha+d)/2}(\kappa \, \|x-y\|)}{\|x-y\|^{(\alpha+d)/2}}  \\ \vspace{3mm}
    & = \frac{2\, (2\kappa)^{(\alpha+d)/2}}{(4\pi)^{d/2}} \J_{(\alpha+d)/2,\kappa}(\|x-y\|),
\end{align*}
with  $$
   \J_{(\alpha+d)/2,\kappa}(s):=\frac{K_{(\alpha+d)/2}(\kappa \, |s|)}{|s|^{(\alpha+d)/2}}, \ s\in \R^*.
$$
Finally, using Fubini's theorem and assuming the convergence of the double integral in \eqref{eq:dbint}, we formally get the following pointwise expression for the operator $-L^{\alpha/2}$, in the Cauchy principal value sense:
$$
    -L^{\alpha/2}v(x)=\tilde{c}_{d,\alpha,\kappa} \int_{\R^d}\J_{(\alpha+d)/2,\kappa}(x-y)(v(y)-v(x)) \, dy -v(x) \, h_\kappa,$$
with $\tilde{c}_{d,\alpha,\kappa}=2\, (2\kappa)^{(\alpha+d)/2}/[(4\pi)^{d/2}\, |\Gamma(-\alpha/2)|]$.

This formula leads to a mechanistic interpretation of the operator $-L^{\alpha/2}$: it contains an absorption term (the reaction term $-v\, h_\kappa$) which becomes stronger as $\kappa$ is increased, and a nonlocal dispersal term of the form \eqref{eq:D_kernel}. The kernel $\J_{(\alpha+d)/2,\kappa}(s)$ behaves as $s^{-(1+\alpha+d)/2}e^{-\kappa \, s}$ for large $s$: it is therefore a thin-tailed kernel, and its tail becomes thinner as $\kappa$ is increased. Contrarily to the fractional Laplacian (with $\alpha\in (0,2)$), and similarly to the standard Laplacian, the operator $-L^{\alpha/2}$ therefore describes short-distance dispersal events. If we come back to the definition of the Mat\'ern covariance \eqref{eq:matern_def}, we note that $\rho_U^{W_S}(h)$ has an exponential decay (like $e^{-\kappa \|h\|}\, \|h\|^{\alpha-d/2-1/2}$) as $\|h\| \to \infty$, corresponding to short-distance dependencies.

\subsubsection{Fractional Laplacian with linear reaction}
The above analysis  shows that the evolving Mat\'ern equation \eqref{eq:evol_Matern} can be written in the form \eqref{eq:convo}, with $D=\tilde{c}_{d,\alpha,\kappa}$, $\J=\J_{(\alpha+d)/2,\kappa}$ and $\kappa^2$ replaced by $h_\kappa$.  This
emphasises the interest of using equations of the general form \eqref{eq:convo} with kernel operators of the form \eqref{eq:D_kernel}  as they should lead to additional flexibility, compared to evolving Mat\'ern equations. We consider here the Eq.~\eqref{eq:convo} with $ D\, (\J\star U- U)(x,t) =-(-\Delta)^{\alpha/2}U(x,t)$, which amounts to considering the fractional Laplacian operator with linear reaction term:
\begin{equation}
-M^{\alpha/2}U := -(-\Delta)^{\alpha/2}U - \kappa^2U,    
\label{eq:M}
\end{equation}
In this case, the (spatial) symbol function $g$ in \eqref{eq:Lg} is $g(\xi)=\|\xi\|^\alpha+\kappa^2$, to be compared with $g(\xi)=(\|\xi\|^2+\kappa^2)^{\alpha/2}$ for the damped Laplacian. The spectral measure is \citep[see Theorem 1 in][]{CarAll21}: $$d\mu_U^W(\xi,\omega)=\frac{1}{(2\, \pi)^{\frac{d+1}{2}}}\frac{d\xi\,d\omega}{\omega^2+(\kappa^2+\|\xi\|^\alpha)^2} \quad \hbox{and} \quad d\mu_U^{W_S}(\xi)=\frac{1}{(2\, \pi)^{\frac{d}{2}}}\frac{d\xi}{(\kappa^2+\|\xi\|^\alpha)^2},$$in the time-dependent and time-independent cases (when $U$ does not depend on $t$, and with a spatial white noise $W_S$), respectively. In the time-independent case, some explicit expressions for the stationary covariance function can be obtained. First, with $\alpha=1$, and in dimension $d=1$, we apply a Fourier transform to obtain the stationary covariance function:
\begin{equation}
\rho_U^{W_S}(h)=\frac{1}{\kappa^2 \, \pi} \left[1- \kappa^2|h|f(\kappa^2|h|)\right], \ h \in \R,
\label{eq:first_aux}
\end{equation}
with $f(z)=\int_0^\infty e^{-z\,t}/(t^2+1) \, dt$. We note that $f$ is the so-called ``first auxiliary function" associated with trigonometric integrals, which can also be written $f(z)= \hbox{Ci}(z) \, \sin (z) + (\pi/2 -\hbox{Si}(z))\cos(z)$  \citep[Chapter 5 in][]{AbrSte64}, with $\hbox{Ci}(z)$ and $\hbox{Si}(z)$ the Cosine and Sine integrals. Using the asymptotic expansions $\hbox{Ci}(z)=\sin(z)/z-\cos(z)/z^2-2\sin(z)/z^3+o(1/z^3)$ and $\pi/2-\hbox{Si}(z)=\cos(z)/z+\sin(z)/z^2-2 \cos(z)/z^3+o(1/z^3)$,  we get  $\rho_U^{W_S}(h)\sim 2/(\kappa^6\,\pi\, |h|^2)$ as $|h|\to \infty$. With the same parameter values ($d=1$, $\alpha=1$), recall that the damped Laplacian leads to the 1/2-Mat\'ern covariance function, which decays exponentially for large $|h|$.

In dimension $d=2$, still with $\alpha=1$, we use the relationship between the Fourier transform of a radially symmetric function and the Hankel transform to compute the stationary covariance:
\begin{align}\label{eq:rho_struve}
\begin{split}
    \rho_U^{W_S}(h) & =\frac{1}{\|h\|^{\frac{d}{2}-1}}\int_0^\infty r^{\frac{d}{2}-1} d\mu_U^{W_S}(r) \, J_{\frac{d}{2}-1}(\|h\| \, r) \, r\,  dr=\frac{1}{2\pi}\int_0^\infty \frac{1}{(\kappa^2 + r)^2} J_{0}(\|h\| \, r)  \, r\, dr \\
    & = \frac{1}{2\pi} \int_0^\infty \frac{\p}{\p (\kappa^2)}\left[\frac{-1}{(\kappa^2 + r)}\right]J_{0}(\|h\| \, r)  \, r\, dr=\frac{1}{2\pi}\frac{\p}{\p (\kappa^2)}\left[ \int_0^\infty \frac{-1}{(\kappa^2 + r)}J_{0}(\|h\| \, r)  \, r\, dr\right],
\end{split}
\end{align}
with $J_\nu$ the Bessel function of the first kind. The last integral in the above equality is a tabulated integral, and corresponds to the Hankel transform of order 0 of $1/(\kappa^2+r)$: from \citep[case 2.12.3.6 in][]{PruBry86} we know that $$\int_0^\infty \frac{1}{(\kappa^2 + r)}J_{0}(\|h\| \, r)  \, r\, dr=\frac{1}{\|h\|} - \frac{\pi \, \kappa^2}{2}\left[(H_{0}-Y_{0})(\kappa^2 \, \|h\|)\right],$$with $Y_{\nu}(\cdot)$ the Bessel function of the second kind and $H_{\nu}(\cdot)$ the Struve function of order $\nu$. Differentiating this expression with respect to $\kappa^2$ and using~\eqref{eq:rho_struve}, we get:
\begin{equation}
\rho_U^{W_S}(h)= \frac{\kappa^2 \, \|h\|}{4} \left[ (Y_{1}-H_{1})(\kappa^2 \, \|h\|)+\frac{2}{\pi}\right]-\frac{1}{4}  \left[(Y_{0}-H_{0})(\kappa^2 \, \|h\|)\right], \ h \in \R^2.
\label{eq:rho_struve_final}
\end{equation}
Here, $\rho_U^{W_S}(h)\sim 1/(\kappa^6 \, \pi \, \|h\|^3)$ as $\|h\| \to \infty$. Note that, with $d=2$ and $\alpha=1$, we have $\alpha=d/2.$ In this case, the damped Laplacian does not lead to a Mat\'ern covariance, but to a generalized covariance function  proportional to $K_0(\kappa\|h\|)\sim e^{-\|h\|}/\sqrt{\|h\|}$ as $\|h\| \to \infty $ \citep{CarAll21,allard2021linking}. 

\begin{figure}
\center
\includegraphics[width=0.6\textwidth]{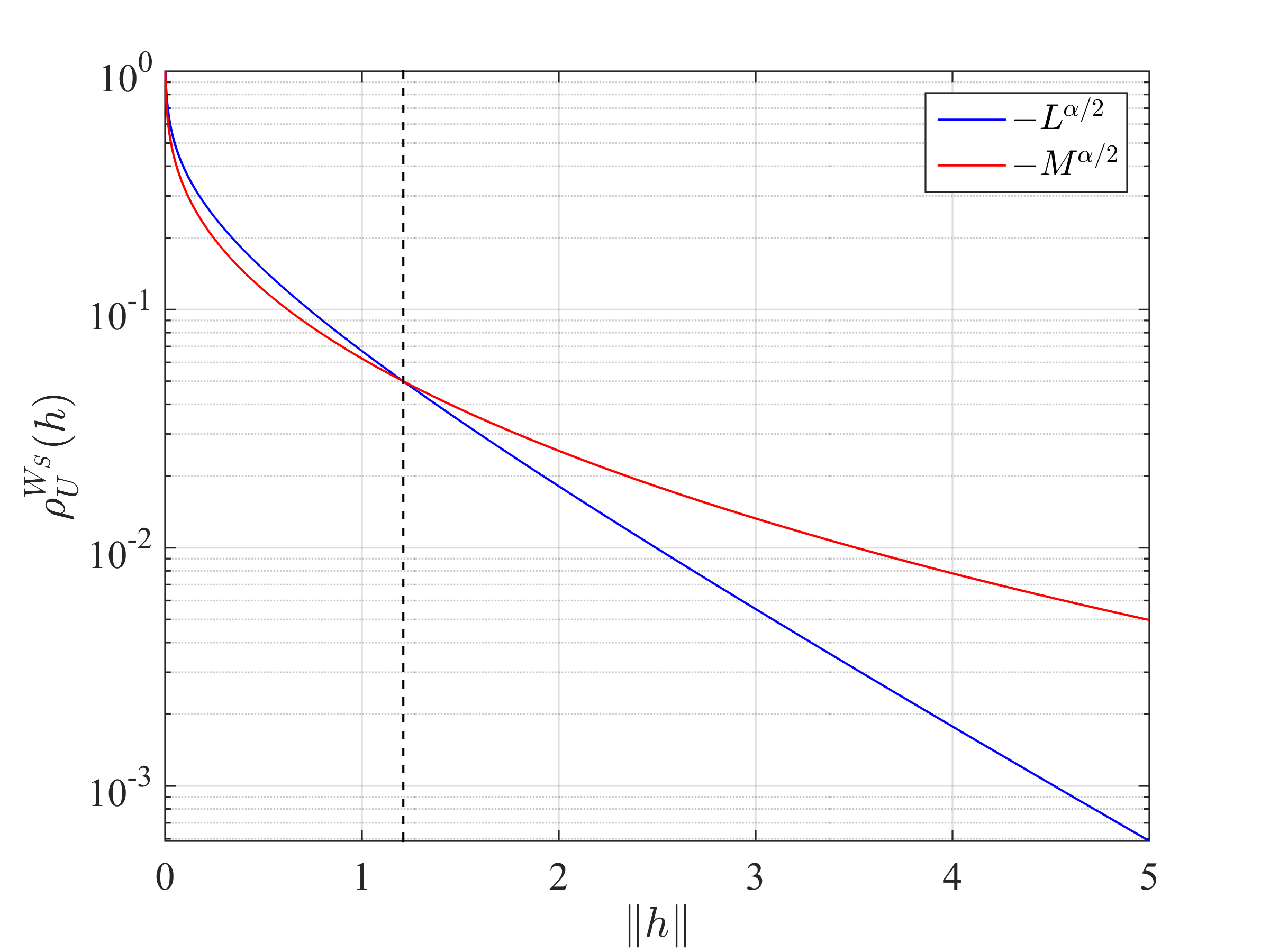}
\caption{{\bf  Covariance functions: damped Laplacian $-L^{\alpha/2}$ vs fractional Laplacian with linear reaction $-M^{\alpha/2}$.} The blue line corresponds to the generalized Mat\'ern covariance function associated with the operator $-L^{\alpha/2}$, and given by \eqref{eq:matern_def}. The red line corresponds to the covariance function associated with the operator~\eqref{eq:M} and given by~\eqref{eq:rho_struve_final}. In both cases, we consider the time-independent problems in dimension $d=2$ and with $\alpha=1$. We set $\kappa=1$ in the operator $-L^{\alpha/2}$, and  $\kappa=0.78$ in the operator $-M^{\alpha/2}$ so that the two covariance functions have same pracical range, i.e. they intersect each other when they are equal to $0.05$ (black dashed line). }
\label{fig:covar}
\end{figure}

In both cases $d=1$ and $d=2$, we have thus observed that the covariance function $\rho_U^{W_S}(h)$ associated with the fractional Laplacian with absorption has an algebraic decay as $\|h\| \to \infty$. As expected from the mechanistic interpretation of the operator, the obtained covariance function therefore describes long-distance dependencies. See Fig.~\ref{fig:covar} for an illustration of the behaviour of the new covariance function \eqref{eq:rho_struve_final}, compared to the generalized covariance given by the damped fractional Laplacian in the case  $d=2$ and $\alpha=1$.  

\subsubsection{Convolution kernels}
We now move away from pseudo-differential operators and consider the equation~\eqref{eq:convo} with regular convolution kernels with mass $1$, for which $\J\star u$ is well-defined in \eqref{eq:D_kernel}. The spectral measures are \citep[Theorem 1 in][]{CarAll21}:
\begin{align*}
    d\mu_U^W(\xi,\omega)&=\frac{1}{(2\, \pi)^{\frac{d+1}{2}}}\frac{d\xi\,d\omega}{\omega^2+(D(1-(2\pi)^{d/2}\F(\J)(\xi))+\kappa^2)^2},\\
    d\mu_U^{W_S}(\xi)&=\frac{1}{(2\, \pi)^{\frac{d}{2}}}\frac{d\xi}{(D(1-(2\pi)^{d/2}\F(\J)(\xi))+\kappa^2)^2},
\end{align*}
in the time-dependent and time-independent cases, respectively. As $\F(\J)(\xi)\to 0$ as $\|\xi\|\to \infty$, these measures are not finite, see Remark~1 in \citep{CarAll21}. In the time-independent case, with an exponential kernel $\J(x)=e^{-\|x\|/\beta}/\left[(2\beta)^d \pi^{\frac{d-1}{2}}\Gamma(\frac{d+1}{2})\right]$, we compute the Fourier transform of $\J$ thanks to the Hankel transform:
\begin{align}\label{eq:fourier_kernel_exp}
\begin{split}
  (2\, \pi)^{d/2}\,\F(\J)(\xi)   & =\frac{(2\, \pi)^{d/2}}{\|\xi\|^{\frac{d}{2}-1}}\int_0^\infty r^{\frac{d}{2}-1} \J(r) \, J_{\frac{d}{2}-1}(\|\xi\| \, r) \, r\,  dr \\ & = \frac{1}{(1+\beta^2\,\|\xi\|^2)^{\frac{d+1}{2}}},
\end{split}
\end{align}
where the last equality follows from \citep[][case 2.8.12.4 with $\nu=d/2-1$ and $\alpha=\nu+2$]{PruBry86}. In the case $d=1$,    
\begin{align*}
d\mu_U^{W_S}(\xi) =&\frac{1}{(2\pi)^{\frac{1}{2}}}\lp\frac{1+\beta^2 \,\xi ^2}{(\kappa^2+D)\beta^2 \xi ^2+\kappa^2}\rp^2  d\xi \\ =&\frac{1}{(2\pi)^{\frac{1}{2}}}\frac{d\xi}{(\kappa^2+D)^2}+\frac{2\, D}{\beta^2 (\kappa^2+D)^3}\frac{1}{(2\pi)^{\frac{1}{2}}}\frac{d\xi}{\xi^2+\kappa^2/[\beta^2(\kappa^2+D)]} \\ & +\frac{ D^2}{\beta^4 (\kappa^2+D)^4}\frac{1}{(2\pi)^{\frac{1}{2}}}\frac{d\xi}{(\xi^2+\kappa^2/[\beta^2(\kappa^2+D)])^2}.
\end{align*}
Since $\F [1/((2\pi)^{\frac{1}{2}} ( \xi^2+a^2))](h)=e^{-a |h|}/(2a)$ and $\F [1/((2\pi)^{\frac{1}{2}} (\xi^2+a^2))^2](h)=(1+ a  |h|)e^{-a |h|}/(4a^3)$ (for $a>0$),
applying a Fourier transform, we obtain an expression for the stationary covariance function:$$\rho_U^{W_S}(h)=\frac{1}{(\kappa^2+D)^2}\delta(h) + (c_{\kappa,D}|h|+\tilde{c}_{\kappa,D})e^{-\frac{\kappa\, |h|}{\beta \sqrt{\kappa^2+D}}}, \ h \in \R$$for some positive constants $c_{\kappa,D}$, $\tilde{c}_{\kappa,D}$. In this case, we observe that the covariance function is comparable to a 3/2-Mat\'ern covariance with an additional Dirac mass at $0$,  referred to as the ``nugget effect" in geostatistics. 

Let us now consider the more general case where the Fourier transform of $\J$ is $ (2\, \pi)^{d/2}\,\F(\J)(\xi) = (1+\beta^2\,\|\xi\|^2)^{-m}$,  with $m\in \N^*$. This corresponds to an exponential kernel with an odd value of the dimension $d=2\,m-1$, see Eq.~\eqref{eq:fourier_kernel_exp} or more generally to a dispersal kernel given by a Matérn covariance in $\R^d$ with parameter $\nu=m$. Then,
\begin{align*}
d\mu_U^{W_S}(\xi) =& \frac{1}{(2\pi)^{\frac{d}{2}}}\frac{(1+\beta^2 \,\|\xi\| ^2)^{2m}}{\left[(\kappa^2+D) (1+\beta^2 \|\xi\| ^2)^{m}-D\right]^2}  d\xi \\
=&\frac{1}{(2\pi)^{\frac{d}{2}}}\frac{d\xi}{(\kappa^2+D)^2} 
+\frac{2\, D}{(\kappa^2+D)^2}\frac{1}{(2\pi)^{\frac{d}{2}}}\frac{d\xi}{(1+\beta^2\|\xi\|^2)^m(\kappa^2+D)-D}\\
& +\frac{ D^2}{(\kappa^2+D)^2}\frac{1}{(2\pi)^{\frac{d}{2}}}\frac{d\xi}{[(1+\beta^2\|\xi\|^2)^m(\kappa^2+D)-D]^2}.
\end{align*}
The  spectral measure associated to a general Matérn dispersal kernel is thus the sum of a Lebesgue measure and two spectral densities which are the inverse of positive polynomials of $\|\xi\|^2$. Thanks to the Rozanov Theorem \citep{rozanov1977markov}, it is known that the Gaussian random fields on $\R^{d}$ characterised by such spectral densities are Gaussian Markov random fields, and thus with thin-tailed covariances.

As the kernel $\J$ were thin-tailed here, the exponential decay of the covariance functions at infinity is not surprising.  We expect a different behavior for fat-tailed kernels \citep[see Table~1 in][]{KleLav06}, leading to long-range dependencies as those observed with the singular kernel $\J(x)=c_{d,\alpha}/\|x\|^{d+\alpha}$ corresponding to the fractional Laplacian.   

\begin{table}[!ht]
    \centering
    \begin{tabular}{lcccc}
     \hline\hline
 Operator  &   Dispersal    & Absorption & Spatial & Covariance  \\
   &  kernel  $\J(\|x\|)$  & term & symbol $g(\xi)$  &  function \\
   \hline
    & & & &\\
\begin{tabular}{@{}l@{}}
                 $-L^{\alpha/2} U$\\
\end{tabular} &  \begin{tabular}{@{}l@{}}
 $\ds \frac{K_{(\alpha+d)/2}(\kappa \, \|x\|)}{\|x\|^{(\alpha+d)/2}}$\\
  $\ds \underset{+\infty}{\sim} c\, \frac{e^{-\kappa \, \|x\|}}{\|x\|^{(1+\alpha+d)/2}} $  \\ $\alpha \in (0,2)$      
\end{tabular}
 &
 \begin{tabular}{@{}l@{}}
 $-h_\kappa \, U$\\ (cf \eqref{eq:hkappa}) \end{tabular}
 & $(\|\xi\|^2+\kappa^2)^{\alpha/2}$  & \begin{tabular}{@{}l@{}}
Matérn \eqref{eq:matern_def}  ($\alpha \ge d/2$) \end{tabular}  \\     
& & & & \\
\hline
& & & & \\
$-M^{\alpha/2}U$  &  $c_{d,\alpha}/\|x\|^{d+\alpha}$  & $-\kappa^2 \, U$ & $\|\xi\|^\alpha+\kappa^2$ &  \begin{tabular}{@{}l@{}}
$d=1,\alpha=1$: \\ \hspace{10mm} \eqref{eq:first_aux} $\ds \underset{+\infty}{\sim}  \frac{2}{\kappa^6\,\pi\, |h|^2} $ \\ $d=2,\alpha=1$: \\ \hspace{10mm}  \eqref{eq:rho_struve_final} $\ds \underset{+\infty}{\sim} \frac{1}{\kappa^6 \, \pi \, \|h\|^3}$\\ $\kappa=0$, $\alpha<d/2$: \\ \hspace{15mm}  $\propto \ds \frac{1}{||h||^{d-2\alpha}}$
                 \end{tabular} \\
& & & & \\
\hline
& & & & \\
\begin{tabular}{@{}l@{}}
                  $D\, (\J\star U- U)$\\\hspace{15mm}$-\kappa^2 U$
\end{tabular}  &  $\J(\|x\|)$  & $-\kappa^2 \, U$ & \begin{tabular}{@{}l@{}}
                  $D(1-(2\pi)^{d/2}\F(\J)(\xi))$\\\hspace{30mm}$+\kappa^2$
\end{tabular}  & \begin{tabular}{@{}l@{}}
                 $\J(x)=e^{-|x|/\beta}$,  $d=1$:\\ ``Nugget"+3/2-Matérn 
\end{tabular} \\ 
  \hline\hline
    \end{tabular}
    \caption{Summary of the main models surveyed in Section~\ref{sec:dispersal}. The (generalized) covariance functions correspond to the time-independent case ($\partial_t U=0$, spatial white noise $W_S$).}
    \label{tab:ccl}
\end{table}

\subsection{Spatially and temporally varying drift terms} 
\label{sec:drift}

The ``heat equation" $\partial_t U(x,t)= D\, \Delta \, U(x,t)$ describes the probability density of a standard Brownian motion. It\^o diffusion generalises Brownian motion by adding a drift coefficient $B(t,X_t)\in \R^d$ (see Section~\ref{sec:Levy}) which takes into account the presence of some external force, or bias. In environmental sciences, such drift may arise from multiple reasons, the effect of the wind on particle movements being one obvious example. The associated density follows an
advection-diffusion equation with a drift term \citep[Fokker-Planck equation; see e.g.][]{Gar09}:
\begin{equation} \label{eq:drift}
    \B (x,t,[U])=-\text{div}(B(x,t)\, U(x,t)),
\end{equation}
where $B=(B_1,\dots,B_d)$ and $\text{div}(B) = \sum_{i=1}^d \partial B_i /\partial x_i$. 
Plugging this drift term into \eqref{eq:SPDE_gale}, with a standard diffusion term $\D(x,t,[U])=D\,\Delta U(x,t)$ and an absorption term $f(x,t,U)=-\kappa^2 \, U(x,t)$, with $D,\kappa>0$, we get the stochastic reaction-advection-diffusion equation 
\begin{equation} \label{eq:SPDE_drift}
\partial_t U(x,t)  = D\,\Delta U(x,t) 
-\text{div}(B(x,t)\, U(x,t))  -\kappa^2 \, U(x,t)+ \sigma W_t(x), \ (x,t)\in \mathbb{R}^d \times \mathbb{R}^+.
\end{equation}
\citet[Section 3.5]{LinRue11} and \citet{sigrist2015stochastic} have considered SPDEs of this form (but with a constant anisotropic diffusion matrix $\mathbf{D}$, which means that $D\,\Delta U(x,t)$ is replaced by $\text{div}( \mathbf{D} \nabla U(x,t))$, see Remark~\ref{rem:aniso}) and constant drift vector $B$, with an application to the modelling of rainfall data.  It is worth emphasising that the drift term \eqref{eq:drift} involves the divergence of the product between the varying drift vector $B(x,t)$ and $U(x,t)$, with $\text{div}(B(x,t)\, U(x,t)) = U(x,t) \text{div}(B(x,t)) + B(x,t)\cdot \nabla U(x,t)$.
When $B$ is constant, as in the above mentioned works, this reduces to $\text{div}(B\, U(x,t)) = B \cdot \nabla U(x,t)$.

\begin{figure}[hbt]
\center
\subfigure[]{\includegraphics[width=0.45\textwidth]{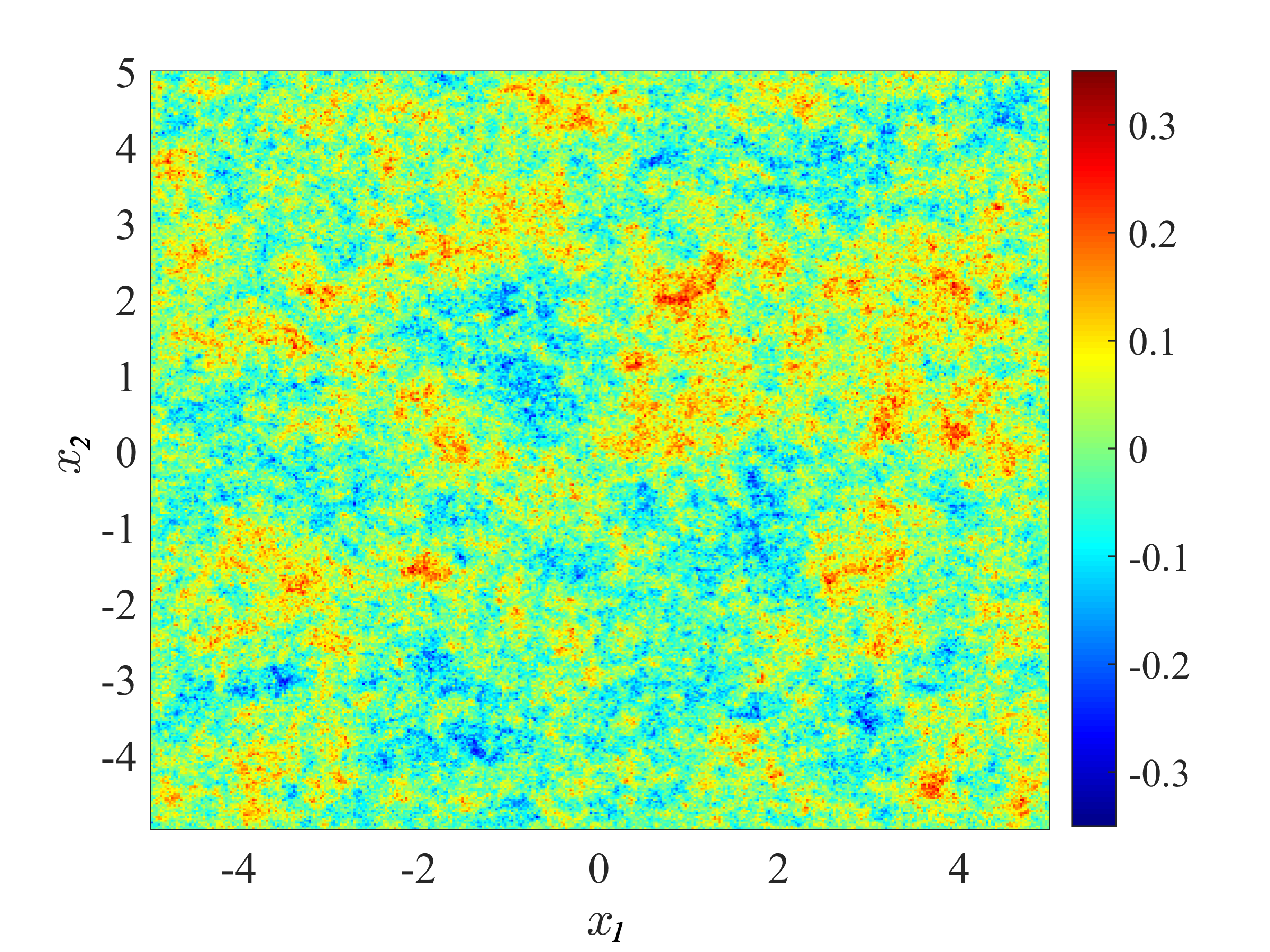}}
\subfigure[]{\includegraphics[width=0.45\textwidth]{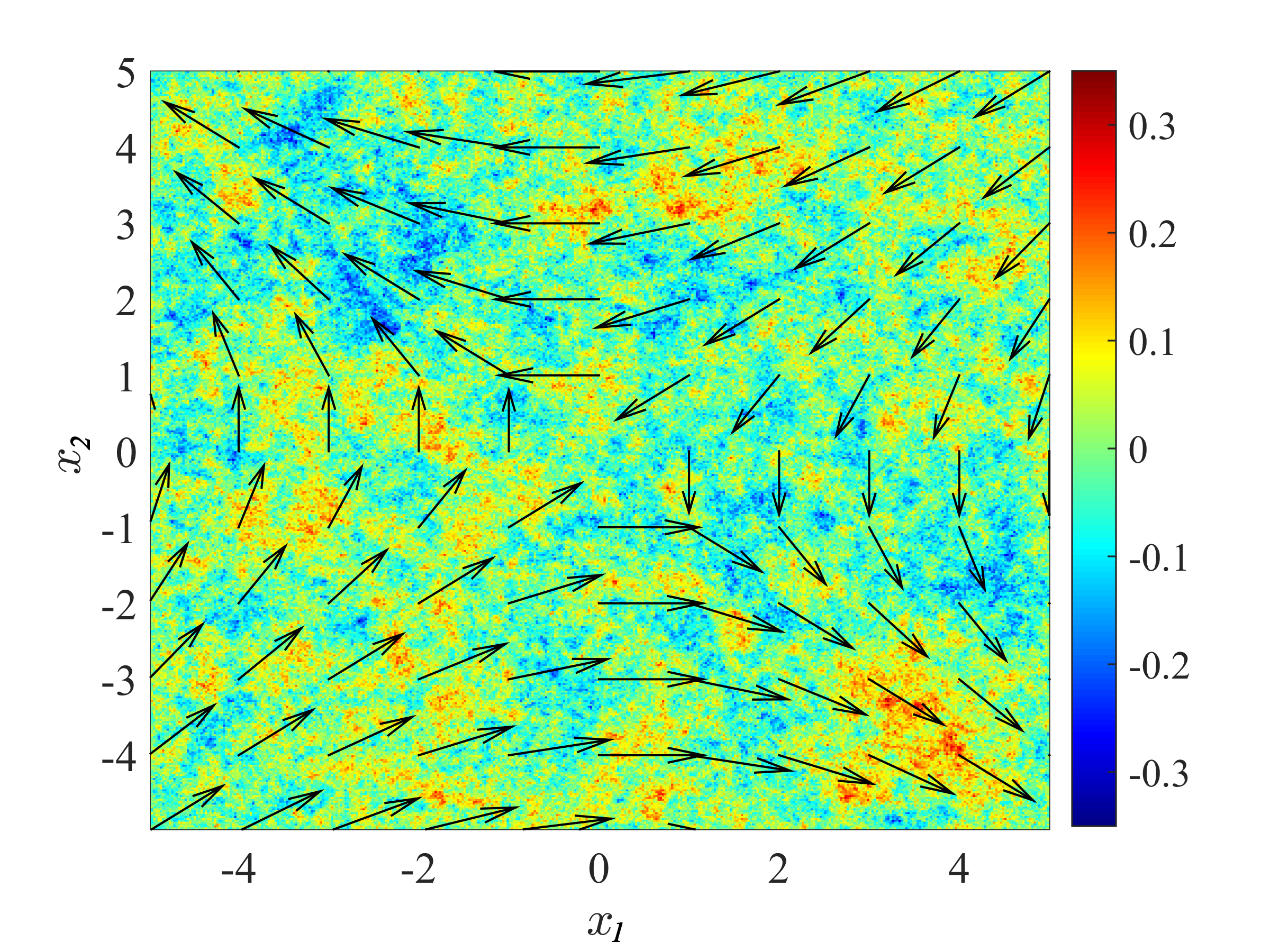}}
\caption{{\bf Effect of a drift term.} Solution of the SPDE \eqref{eq:SPDE_drift} with null drift coefficient $B(x,t)\equiv 0$ in panel (a), and a drift coefficient $B(x,t)$ given by \eqref{eq:driftB} in panel (b). In both cases, $D=0.05$, $\kappa=0.1$, $\sigma=1$ and the solution is plotted at a fixed time $T=6$. In panel (b), the arrows correspond to the directional components of the vector field $B$.}
\label{fig:drift}
\end{figure}

To the best of our knowledge, a drift vector that varies in space and time has never been considered. It is beyond the scope of this paper to mathematically characterise the associated random fields, as the operator does not fit in the class considered in \citet{CarAll21} when the vector field $B$ is not constant.  In Fig.~\ref{fig:drift}, we compare the solution of equation \eqref{eq:SPDE_drift} in a square domain of $\R^2$ at a fixed time $T$, with the solution of the standard equation~\eqref{eq:basic} without drift term. In both cases, we start from an initial condition $U(x,0)=0$, and we approach the solution with a finite difference approximation over a bounded domain with Dirichlet boundary conditions, i.e. $U(x,t)=0$ on the boundary of the domain. We  use an explicit Euler scheme in time (Matlab codes are available in the Open Science Framework repository: \url{https://osf.io/w5utd}). The drift term is chosen as follows:
\begin{equation}
    \left\{
    \begin{array}{l}
         B_1(x_1,x_2,t)= 2\, x_2 \cos(2\, \pi \, t/10)/\sqrt{0.1+x_1^2+x_2^2}, \\
         B_2(x_1,x_2,t)= 2\, x_1 \sin(2\, \pi \, t/10)/\sqrt{0.1+x_1^2+x_2^2}.
    \end{array}
    \right.
    \label{eq:driftB}
\end{equation}
The other parameter values are given in the legend of Fig.~\ref{fig:drift}. The effect of the drift term is more visible in the video file \url{https://osf.io/ydrqk/}, to be compared with the dynamics in the absence of drift \url{https://osf.io/krvug/}.

% % Y=sol_EDPS_Euler_basic(0.05,0.05,0.1,1,10,10,10,1);
% Y=sol_EDPS_Euler_drift(0.05,0.05,0.1,1,10,10,10,1);
% 

\subsection{Bistable reaction terms} 
\label{sec:bistable}
In the absence of noise, that is when $\sigma=0$, $u(x,t)=0$ is the only solution of \eqref{eq:basic} as $t \to \infty$  and it is asymptotically stable (notice that we use here the lower case $u$ to denote the solution since it is not random). In other words, in the PDE framework, starting from any bounded initial condition, the solution $u(x,t)$ converges towards $0$ uniformly in space and time. This easily follows from a comparison argument \citep{ProWei67}: $\min_{x\in \R^d} u(x,0) \, e^{-\kappa^2 \, t} \le u(x,t) \le \max_{x\in \R^d} u(x,0) \, e^{-\kappa^2 \, t}$ for all $t\ge 0$ and $x\in \R^d$. On the other hand, if the reaction term $-\kappa^2 \, u(x,t)$ is replaced by a bistable reaction term:
\begin{equation} \label{eq:bistable}
f(x,t,u)=u(x,t) \, (K-u(x,t)) \, (u(x,t)-\rho),
\end{equation}
with $K>0$ and $\rho \in (0,K)$, the equation exhibits much more complex dynamics. Ignoring dispersal and drift, the ordinary differential equation $v'(t)=f(v)$, $t>0$ admits two stable stationary states: $0$ and $K$, and an unstable one: $\rho$. Thus, starting with an initial condition $v(0) \in (-\infty,\rho)$, $v(t)$ ultimately converges to $0$, and if $v_0 \in (\rho,+\infty)$, $v(t)$ converges to $K$ \citep[see e.g.][]{Roq13}. In the PDE framework, when $u(x,t)$ also depends explicitly on $x$, the behaviour of the solution strongly depends on the precise shape of the initial condition $u(\cdot,0)$ and on the value of $\rho$. This is the consequence of the interplay between the dispersal and the reaction term. Consider for example $u(x,0)= K \, \chi_\Omega(x)$, where  $\chi_\Omega$ is the characteristic function of some set $\Omega \subset \R^d$. Then the solution tends to converge towards $K$ for large and aggregated sets $\Omega$ and tends to converge towards $0$ for small and fragmented sets $\Omega$ \citep{Zla06,DuMat10,GarRoqHam12}. More generally, the sign of the integral of $f$ over $(0,K)$ determines which state is more stable: $0$ if the integral is nonpositive and $K$ if the integral is positive. The ``more stable" state tends to invade the other one for a larger class of initial conditions \citep[e.g.,][]{FifMcL77}. The particular case $\rho=K/2$ corresponds to a fully symmetric situation.

The reaction term~\eqref{eq:bistable} is often used in biology to take into account an ``Allee effect" \citep{LewKar93,KeiLew01}. 
In such case $f(x,t,u)$ corresponds to the growth rate of a population with local density $u(x,t) \ge 0$. The local population density must be above the ``Allee threshold" $\rho$ to reach a positive growth rate. This means that a form of cooperation between the individuals is required. When the population density is too large ($u(x,t)>K$) the growth rate becomes negative again, meaning that the population density tends to decay due to e.g. resource saturation when it reaches too large values. From a more empirical viewpoint, this reaction term reproduces the bistable behaviour that can be observed in some ecosystems. For instance, it is known that a positive feedback between the vegetation abundance and the local conditions for the growth of other plants leads to bistability and patchiness, especially in arid systems \citep{RieKop97,KefRie07}.

\begin{figure}
\center
\subfigure[$D=0.01$]{\includegraphics[width=0.45\textwidth]{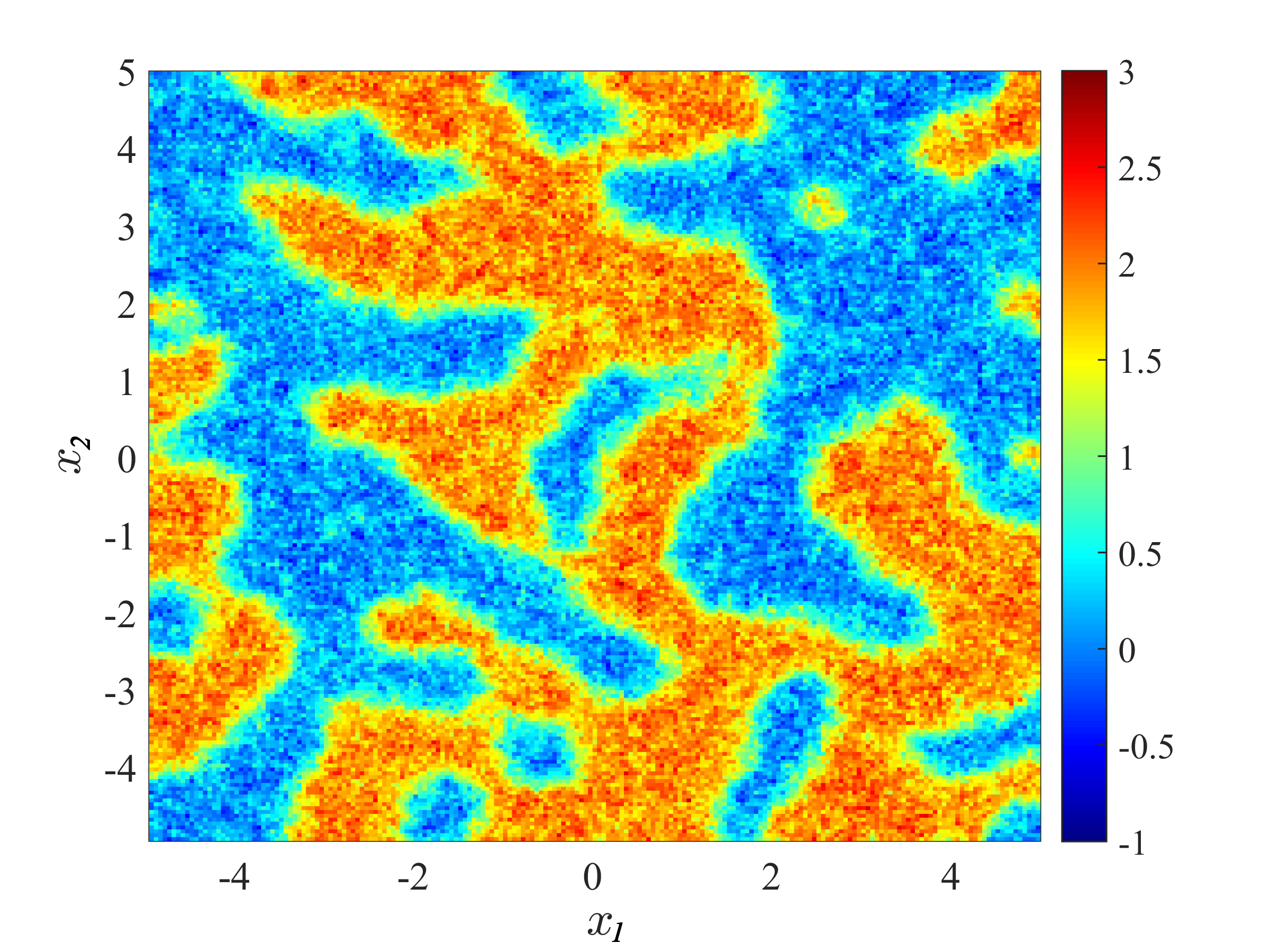}}
\subfigure[$D=0.1$]{\includegraphics[width=0.45\textwidth]{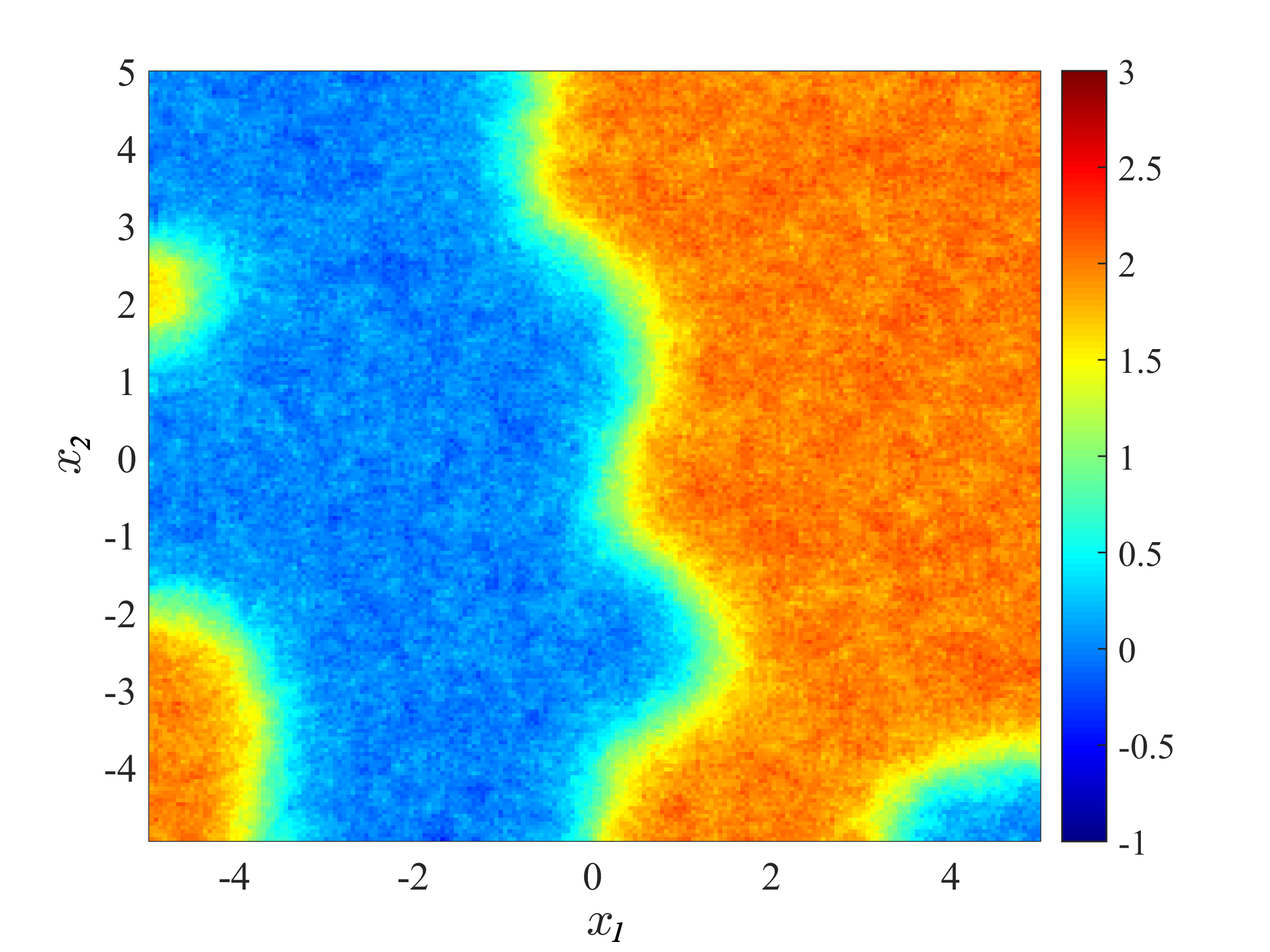}}
\subfigure[$D=0.01$]{\includegraphics[width=0.45\textwidth]{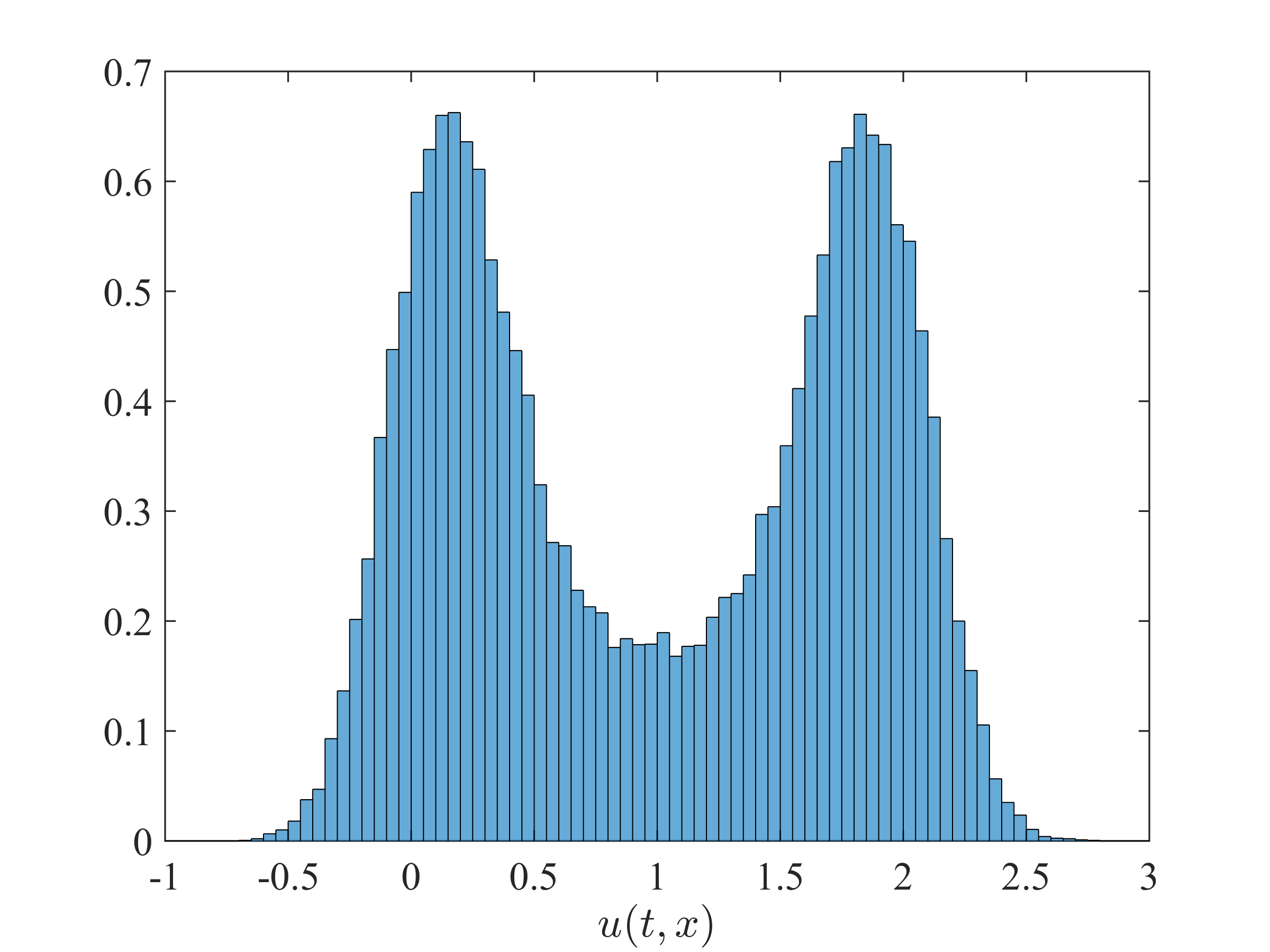}}
\subfigure[$D=0.1$]{\includegraphics[width=0.45\textwidth]{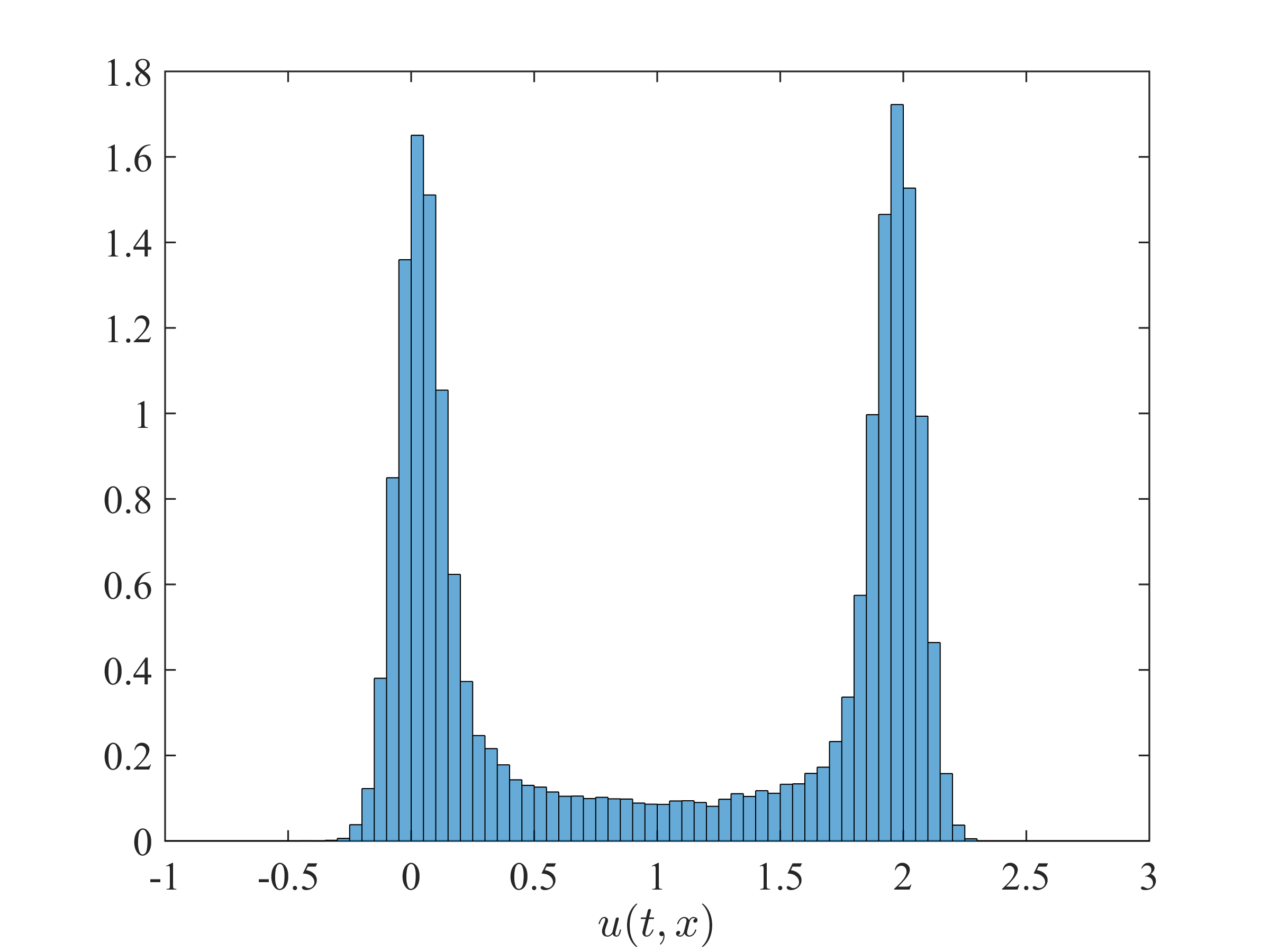}}
\caption{{\bf Effect of a bistable reaction term.} Panels (a,b): finite difference solution of the SPDE \eqref{eq:Allen-Cahn}. Panels (c,d): corresponding histograms describing the distribution of $U(x,T)$. In all cases, $K=2$, $\sigma=1$ and $T=10$. The mesh size is $\delta x=0.05$. }
\label{fig:bistable}
\end{figure}

Here, we consider the equation \eqref{eq:SPDE_gale} with a diffusive dispersal term $\D(x,t,[U])=D\,\Delta U$, a null drift term $\B (x,t,[U])=0$ and a reaction term \eqref{eq:bistable} with $\rho=K/2$. This corresponds to the stochastic Allen-Cahn equation, 
\begin{equation}\label{eq:Allen-Cahn}
    \p_t U(x,t)= D\,\Delta U(x,t)  -\kappa^2 \, U (x,t) + U(x,t) \, (K-U(x,t)) \, (U(x,t)-K/2)\sigma\,  W_t(x),
\end{equation}
for which existence and uniqueness are only established when the space dimension is $d=1$. When $d\ge2$, as in the examples that we consider in this work, it is conjectured that the equation is ill-posed, and it was shown that modelling with this equation is questionable, because the numerical solutions converge to the zero-distribution when the mesh grid is refined \citep{RysNig12}. Several studies work around this problem by considering a coloured noise with a finite spatial correlation length $\lambda$ \citep[e.g.][]{KohOtt07}. In spite of these difficulties, and even though the theory in \cite{CarAll21} does not apply in this case due to the nonlinear nature of $f$, we foresee that the numerical solution of the SPDE may exhibit interesting properties, especially to model variables with bimodal distributions. As argued by \cite{RysNig12}, although the SPDE \eqref{eq:Allen-Cahn} may be ill-posed, it is still possible to give an interpretation to its ``finite difference solution", with a fixed mesh size $\delta x$: it shall be interpreted as the solution of an equation driven by a noise with a finite correlation length $\lambda$ which is much smaller than $\delta x$.  We provide an example in Fig.~\ref{fig:bistable}. Here we start from an initial condition $U(x,0)=K/2$ for $x \in [-5,5] \times [5,5]$. Using an explicit Euler scheme in time, we compute the finite difference approximation with a regular mesh size $\delta x=0.05$, over a bounded domain with Neumann boundary conditions, i.e. $\nabla U(x,t)=0$ when $x$ is on the boundary of the domain. Matlab codes are available at \url{https://osf.io/w5utd/}.
We observe in panels (a,b) that the corresponding random fields tends to be much more clustered than those generated by \eqref{eq:basic} (compare with Fig.~\ref{fig:drift}a). Moreover, increasing  diffusion coefficients $D$ tend to create larger clusters. As expected from the discussion above, at a fixed time $T$, $U(x,T)$ has a very clear bimodal distribution despite the Gaussian random noise $W_t$, with one mode close to 0 and the other one close to $K$ (see Figs.~\ref{fig:drift}c and \ref{fig:drift}d).

%Y=sol_EDPS_Euler_bistable(0.1,1,2,1,1,10,10,10,1);
%Y=sol_EDPS_Euler_bistable(0.01,1,2,1,1,10,10,10,1);

\

\section{SDEs and point processes: the case of dispersing particle systems}
\label{sec:PDEs}

Here, we describe a theoretical framework for modelling the spatio-temporal dynamics of particles and explore the point processes that such a framework generates. The point process viewpoint is then used to tackle the estimation of model parameters using spatio-temporal observations. This estimation task naturally fits into the mechanistic-statistical approach \citep[e.g.,][]{SouRoq14}, which bridges the gap between mechanistic models (here a L\'evy flight model describing particle dynamics; see Eq. \eqref{eq:levydiffusion}) and data, via a probabilistic observation model. Here, observations are derived from an inhomogeneous spatio-temporal point process whose intensity follows a PDE, namely the Fokker-Planck equation \eqref{eq:FP_Levy_u} associated with the stochastic dispersal process.

\paragraph{A stochastic model for the dispersal and deposition of particles.} The positions of the particles at any time $t\ge 0$ are described by a vector field $\mathbf{X}_t$ of size $d \times N_t$, where $d$ is the space dimension, usually $2$ or $3$, and $N_t$ is the number of particles in our system at time $t\ge 0$. 
The initial number of particles $N_0$ follows a Poisson distribution with parameter $E[N_0] = \eta_0$, and these particles are distributed in $\R^d$ according to a probability distribution function $p_0$ \citep[hence, the initial pattern of particle locations is drawn from an inhomogeneous Poisson point process;][]{illian2008,chiu2013}.  Between $t=0$ and the deposition times, the particle dynamics are described by independent $d-$dimensional $\alpha-$stable L\'evy processes, for some $\alpha\in (0,2]$, see Eq.~\eqref{eq:levydiffusion}.

As in Section~\ref{sec:Levy}, we assume that the expected duration of the dispersal before deposition (or removal) is  $1/\kappa^2>0.$ The deposition events are independent and identically distributed from an exponential distribution with parameter $\kappa^2$. Namely, for each $i\in \{1, \ldots,N_0\},$ the deposition time $\tau_i$ of the particle $i$ satisfies $\tau_i \sim \hbox{Exp}(\kappa^2).$

\paragraph{Fokker-Planck equations.} As already mentioned in Section~\ref{sec:Levy}, the expected density of still dispersing particles $u(x,t)$ satisfies Eq.~\eqref{eq:FP_Levy_u}. Here, $N_0$ is a random variable, and the initial condition is 
$u_0(x):=u(x,0)=\eta_0\, p_0(x).$
The only non-conservative term in Eq.~\eqref{eq:FP_Levy_u} satisfied by $u$ is $- \kappa^2 \, u$. It corresponds to particle removal due to deposition. The expected density $w(x,t)$ of deposited particles thus satisfies $\partial_t w(x,t)=\kappa^2 \, u(x,t), \ t>0, \ x\in\R^d,$ with initial condition $w(\cdot,0)=0$ (all of the particles are dispersing at the initial time $t=0$).

\paragraph{Inhomogeneous spatial Poisson point process.} At a fixed time $t$, the spatial point process formed by the dispersing particles is an inhomogeneous Poisson point process with intensity function $u(\cdot,t)$; see a more precise formulation of this rather straightforward result in Appendix~A, Lemma~\ref{lem:poisson}. Indeed, let $A$ be a bounded region in $\R^d$, and denote $N_t(A)$ the number of dispersing particles in $A$ at time $t$. We have
$$\pr(N_t(A)=m)=\sum_{k=m}^\infty \pr(N_t(A)=m|N_0=k) \pr(N_0=k).$$
As the particles follow independent movements, $$\pr(N_t(A)=m|N_0=k)=  \binom{k}{m}\rho^m (1-\rho)^{k-m},$$ with $\rho=\int_A h(y,t)\, dy$ the probability that a given particle is still dispersing (alive) and belongs to $A$ at time $t$. The function $h$ was defined in Section \ref{sec:Levy}; it satisfies the same equation as $u$ but with the initial condition $h(\cdot,0)=p_0$. Thus, $$\pr(N_t(A)=m)=\sum_{k=0}^\infty \binom{k}{m}\rho^m (1-\rho)^{k-m}\frac{\eta_0^k}{k!} e^{-\eta_0}=\frac{\rho\, \eta_0^m}{m!}e^{\rho\eta_0}.$$As $\rho\, \eta_0=\int_A u(y,t)\, dy,$ this shows that,  at each fixed time $t$, the distribution of dispersing particles follows an inhomogeneous Poisson point process with intensity $u(\cdot,t)$.
The same reasoning implies that, at each fixed time $t$, the distribution of deposited (dead) particles follows an inhomogeneous Poisson point process with intensity $w(\cdot,t)$.

%We use these properties to model the observations and compute a likelihood.

\paragraph{Non-Poisson spatio-temporal point process.}
Despite the independence of particle trajectories, the spatio-temporal point process formed by the locations of still-dispersing particles is not a spatio-temporal Poisson point process because of the temporal dependencies of the successive locations of any particle (see Appendix~A, Lemma~\ref{lem:dependence} with $A=\R^d$). However, we can state that the point process $\X_t$ formed by the particle locations at time $t$ given the particle locations $\X_s$ ($s<t$) is the union of  $N_s$ independent thinned binomial point processes with size 1, thinning probability $1-\exp(-(t-s)\kappa^2)$, and p.d.f.  $p_{\delta(\cdot-X_s^i)}^s(\cdot,t)$, where
$X_s^i$ is the point of $\X_s$ with label $i$ and $p_{X_s^i}^s$ satisfies Eq.~\eqref{eq:FP_Levy} with initial condition $p_{X_s^i}^s(\cdot,s)=\delta(\cdot-X_s^i)$ at time s (see Appendix~A, Lemma~\ref{lem:union} with $A=\R^d$).
More generally, based on Lemma~\ref{lem:union}, the point process  $\X_t$ conditional on particle locations in a subset $A\subset \R^d$ at time $s<t$ (without being able to {\it recognise} the particles and {\it follow them} across sampling times) leads to a point process made of the union of an inhomogeneous Poisson point process and independent thinned binomial point processes with size 1. 

\paragraph{Likelihood-- and pseudo-likelihood--based estimation.} Consider a family of observation windows $\{A_k\}_{k=1,\ldots,K}$, and an increasing sequence $\{t_k\}_{k=1,\ldots,K}$ of observation times. We focus on the computation of a likelihood based on the observation of dispersing particles located within the spatial$\times$time windows $\{A_k,t_k\}_{k=1,\ldots,K}$ to estimate the model parameters; we could use observations of deposited particles as well. 

Let $\X^{(k)}$ denote the observation of $\X_{t_k}$ restricted to $A_k$. The likelihood satisfies 
\begin{equation*}
\Ll(\Theta):=\pr(\X^{(1)},\ldots,\X^{(K)})= \pr(\X^{(1)}) \prod_{k=2}^K \pr(\X^{(k)}\mid\X^{(1)},\ldots,\X^{(k-1)}),
\end{equation*}
where $\Theta$ is the vector of model parameters to be estimated.
The conditional probability $\pr(\X^{(k)}\mid\X^{(1)},\ldots,\X^{(k-1)})$ simplifies into $\pr(\X^{(k)}\mid\X^{(k-1)})$ only if $A_{k-1}=\R^d$ (in this case, for any $k'\ge k$, $\pr(\X^{(k')}\mid\X^{(1)},\ldots,\X^{(k'-1)})=\pr(\X^{(k')}\mid\X^{(k-1)},\ldots,\X^{(k'-1)})$). If $A_{k-1}\neq\R^d$ for some $k$ (a very standard situation), the likelihood can be written as a multiple integral with respect to the unobserved parts of $\X_{t_1},\ldots,\X_{t_{K-1}}$, say $\tilde\X^{(1)},\ldots,\tilde\X^{(K-1)}$ (with $\tilde\X^{(k)}=\X_{t_k} \backslash \X^{(k)}$). Namely, we show by induction (see Appendix~A.3) that: 
\begin{align}\label{eq:likelihood}
\Ll(\Theta)&= \int_{\tilde\X^{(K-1)},\ldots,\tilde\X^{(1)}}
\pr(\X^{(K)}\mid \X_{t_{K-1}})
\left(\prod_{k=2}^{K-1}d\pr(\X_{t_{k}}\mid\X_{t_{k-1}})\right)
d\pr(\X_{t_1}).
\end{align}
$\pr(\X_{t_1})$ is the probability measure of the point pattern $\X_{t_1}$, which is drawn from an inhomogeneous Poisson point process with intensity $u(\cdot,t_1)$ satisfying Equation \eqref{eq:FP_Levy_u} with initial condition $\eta_0 p_0$ at $t=0$.
$\pr(\X_{t_{k}}\mid\X_{t_{k-1}})$ can be written:
\begin{equation}
\label{eq:XgivenXm1}
\begin{split}\pr(\X_{t_{k}}\mid\X_{t_{k-1}})
&= \sum_{\psi_k} \pr(\X_{t_{k}}\mid\X_{t_{k-1}},\psi_k)
\pr(\psi_k)\\
&= \sum_{\psi_k}\left( \prod_{i=1}^{N_{t_k}} p_{\delta(\cdot -X_{t_{k-1}}^{\psi_{ki}})}^{t_{k-1}}(X_{t_k}^i,t_k)\right) \pr(\psi_k),
\end{split}
\end{equation}
where the sum is over all the ordered samplings $\psi_k$ of $N_{t_k}$ elements from $N_{t_{k-1}}$ elements (without replacement). Namely, $\psi_k=(\psi_{k1},\ldots,\psi_{kN_{t_k}})$ where $\psi_{ki}$ indicates to which point in $\X_{t_{k-1}}$ corresponds the $i$-th point $X_{t_k}^i$  in $\X_{t_k}$ 
and $p_{\delta(\cdot -X_{t_{k-1}}^{\psi_{ki}})}^{t_{k-1}}(\cdot,t_k)$ (defined in the notation paragraph of Appendix A) is the probability density function of $X_{t_k}^i$ given the particle $i$ was at $X_{t_{k-1}}^{\psi_{ki}}$ at time $t_{k-1}$. Note that $N_{t_k}$ is finite since it is less than $N_0$, which is Poisson-distributed with mean $\eta_0$. Similarly to $\pr(\X_{t_{k}}\mid\X_{t_{k-1}})$, $\pr(\X^{(K)}\mid \X_{t_{K-1}})$ can be written:
\begin{equation}
\label{eq:XKgivenXKm1}
\pr(\X^{(K)}\mid\X_{t_{K-1}})
= \sum_{\psi_K}\left( \prod_{i=1}^{N_{t_K}(A_K)} p_{\delta(\cdot -X_{t_{K-1}}^{\psi_{Ki}})}^{t_{K-1}}(X^{(K)i},t_K)\right) \pr(\psi_K),
\end{equation}
where $X^{(K)i}$ is the location of the $i$-th point of $\X^{(K)}$.

Maximum likelihood estimation or Bayesian estimation can be performed using the likelihood expression \eqref{eq:likelihood}. The multiple integration (with respect to the latent point patterns $\tilde\X^{(1)},\ldots,\tilde\X^{(K-1)}$ over $A_1,\ldots,A_{K-1}$ at times $t_1,\ldots,t_{K-1}$ and the latent vectors $\psi_2,\ldots,\psi_K$) may be handled with stochastic estimation algorithms such as SEM, MCEM, MCMC or adaptive importance sampling. However, such approaches likely require high computational cost.

Alternatively, we propose here to use a pseudo-likelihood function constructed by ignoring the temporal dependencies except the dependencies generated by the temporal evolution of the intensity function of $\X_t$, that is to say the intensity function $u$ satisfying Equation \eqref{eq:FP_Levy_u} with initial condition $\eta_0 p_0$; see Appendix A, Lemma \ref{lem:poisson} and Corollary \ref{cor:intensity}.
Thus, we ignore dependencies between observations at successive observation times and make as if  $\{\X_t\}_{t\ge 0}$ is an inhomogeneous spatio-temporal Poisson point process with spatio-temporal intensity $u$ ($u$ is thereafter denoted by $u_\Theta$ to highlight the dependence of $u$ with respect to the parameters). 
%As explained above, we assume that  the dispersing particles have an inhomogeneous Poisson distribution with intensity $u(\cdot,t)$ in each observation window and at each fixed time. Assuming that the observations are independent from each other conditionally on the intensity, we obtain the likelihood \citep[see e.g.][]{MolWaa07}:
Consequently, the pseudo-likelihood satisfies \citep[see e.g.][]{MolWaa07}:
\begin{equation}
    \label{eq:like1}
\pLl(\Theta)=\prod\limits_{k=1}^K e^{|A_k|-\varphi_{\Theta,t_k}(A_k)}\prod\limits_{X_{t_k}^i\in A_k} u_\Theta(t_k,X_{t_k}^i),
\end{equation}
with $|A_k|$ the Lebesgue measure of $A_k$, and $\ds \varphi_{\Theta,t_k}(A_k)=\int_{A_k} u_\Theta(t_k,Y)\, dY$ the intensity measure of $A_k$ at time $t_k$.
%and $u_\Theta$ the solution of \eqref{eq:FP_Levy_u}, with parameter $\Theta$.

Assume now as a benchmark that the observations only consist of counting data, i.e., numbers of particles in windows $A_k$ at times $t_k$, $k=1,\ldots,K$. Ignoring as above temporal dependencies except the dependencies generated by the temporal evolution of the intensity function, the numbers of particles in windows $A_k$s at observation times $t_k$s follow independent Poisson distributions with mean values $\varphi_{\Theta,t_k}(A_k)$ (intensity measure of $A_k$), and the pseudo-likelihood now reads:
\begin{equation}
\label{eq:like_count}
\pLl_c(\Theta)=\prod\limits_{k=1}^K[\varphi_{\Theta,t_k}(A_k)]^{N_{t_k}(A_k)}\, \frac{e^{-\varphi_{\Theta,t_k}(A_k)}}{N_{t_k}(A_k)!}.
\end{equation}
%with $N(A_k)$ the number of observed particles in $A_k$ at time $t_k$. 

\begin{figure}
\center
\subfigure[$t_1=1$]{\includegraphics[width=0.32\textwidth]{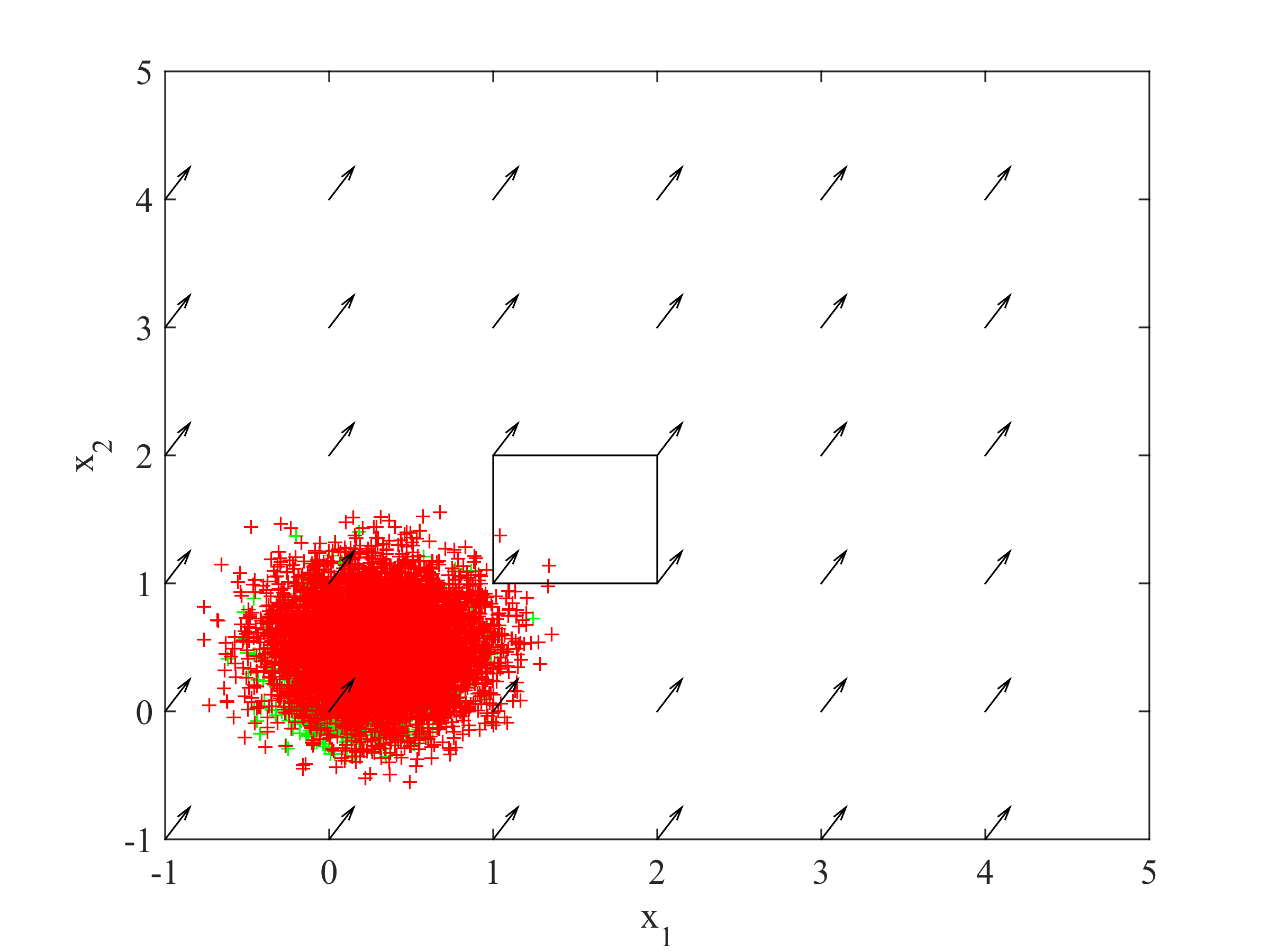}}
\subfigure[$t_2=3$]{\includegraphics[width=0.32\textwidth]{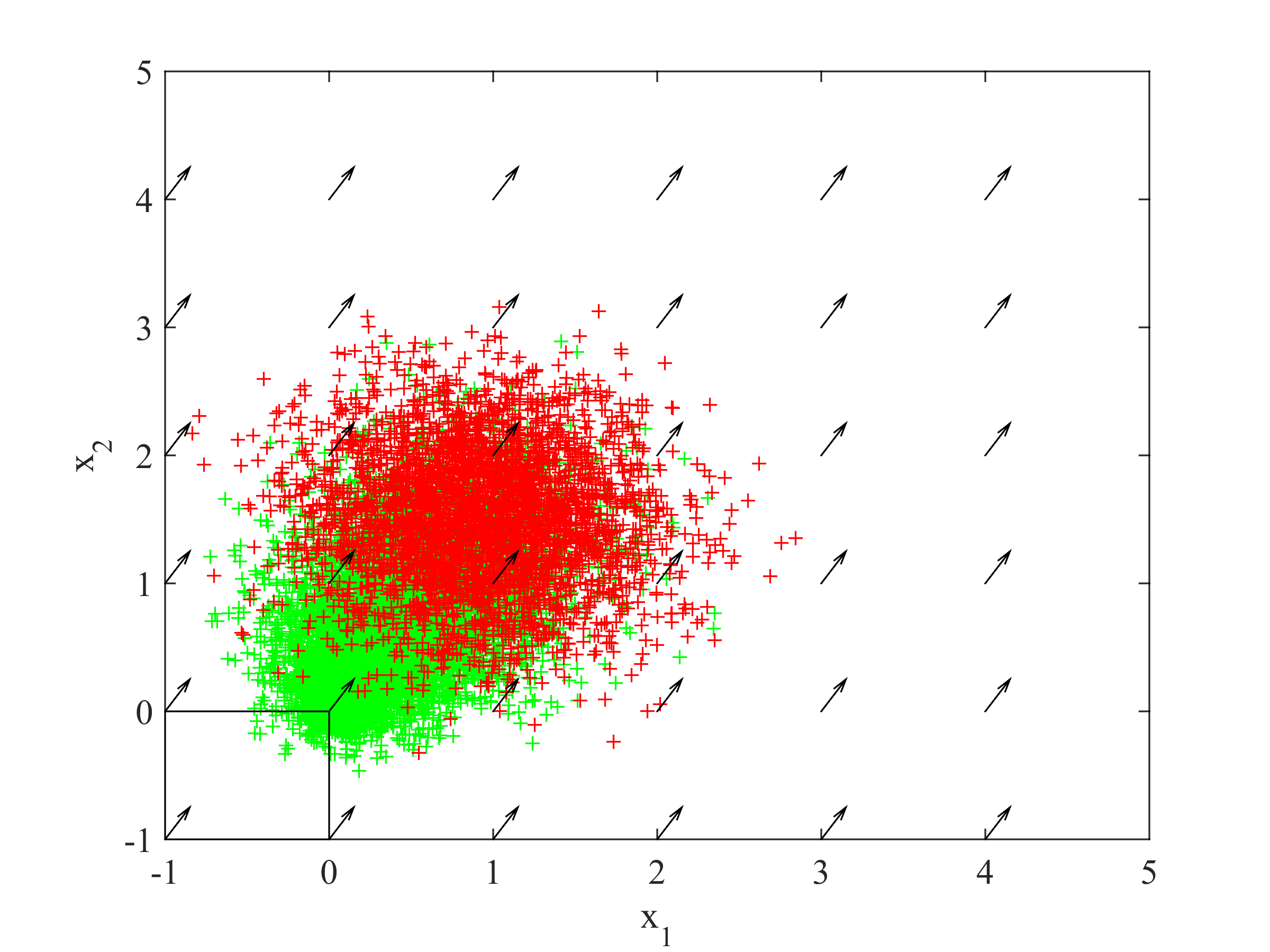}}
\subfigure[$t_3=6$]{\includegraphics[width=0.32\textwidth]{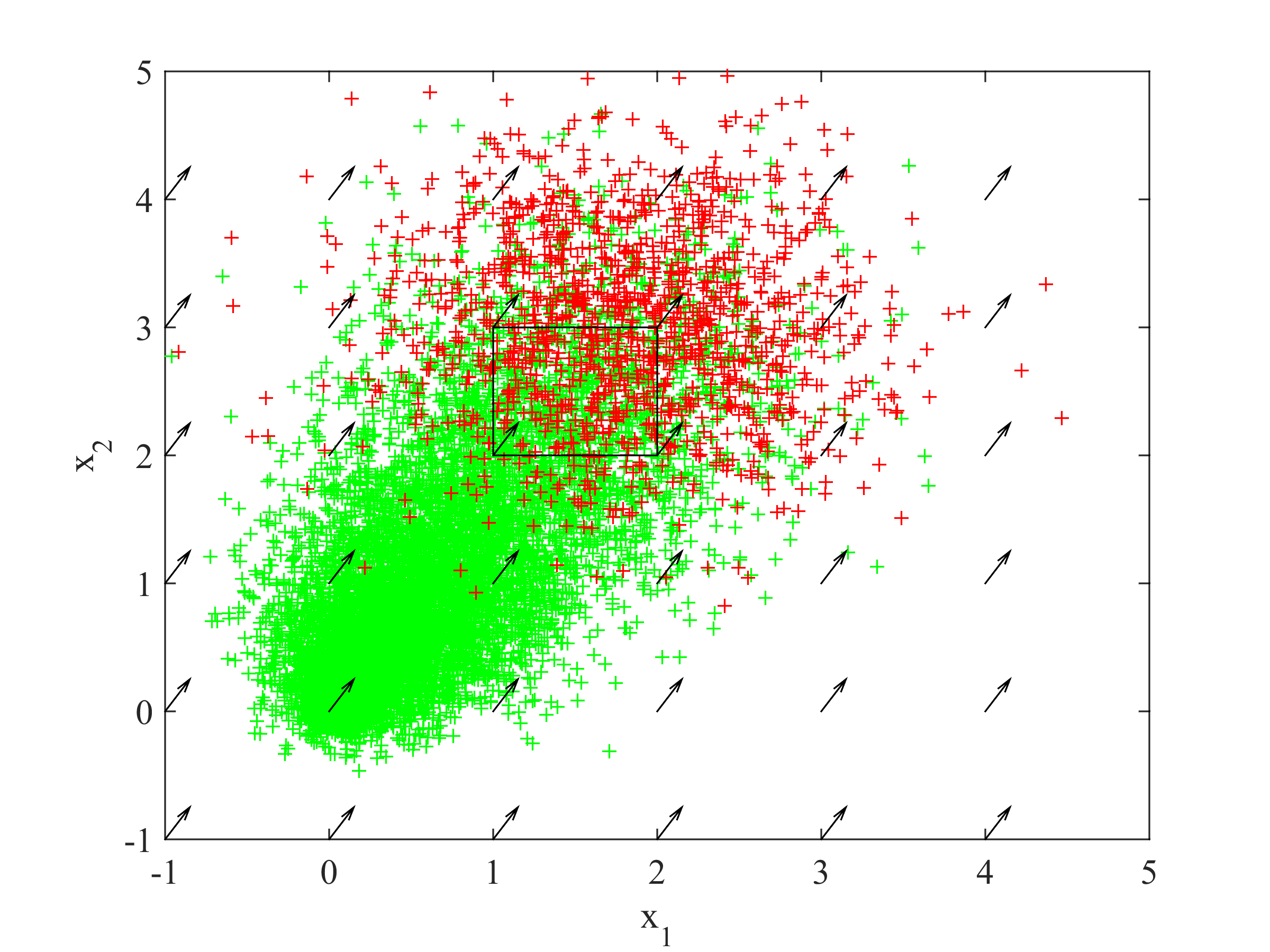}}
\caption{{\bf Observation windows.} Dispersing particles (red crosses) and deposited particles (green crosses) at successive times $t=1,\, 3, \,6$.  The black arrows describe the vector field $B$, and the black rectangles are the observation windows $A_1$, $A_2$, $A_3$. The particles were initially uniformly distributed in a disc of radius $0.05$, centered at $0$.}
\label{fig:particles}
\end{figure}

Thus, in the proposed mechanistic-statistical approach, we consider a unique mechanistic model (for which we aim to estimate parameters) but we consider two different  observation processes, consisting of either observing particle locations in $A_k$s at $t_k$s (in which case $\pLl(\Theta)$ is maximised), or counting particles in $A_k$s at $t_k$s, (in which case $\pLl_c(\Theta)$ is maximised).  
\paragraph{Numerical example.} We ran $10^3$ simulations of the discrete particle system in $\mathbb{R}^2$ with a Brownian diffusion ($\alpha=2$, $\gamma=1/2$),  and spatially-homogeneous coefficients $B=(0.3,0.5)$, $\sigma=0.3$ and mean deposition time $1/\kappa^2=3$. The particles were released in a small disc around $x=0$, with an initial number drawn in a Poisson distribution with parameter $\eta_0=10^4$. Assuming that the initial distribution is $p_0(x)=\eta_0\,\delta(x)$, the solution $u(x,t)$ of the Fokker-Planck equation has an analytic expression 
\begin{equation}
\label{eq:expr_u_homo4}
\ds \tilde u (x,t)=\frac{\eta_0\,  e^{-\kappa^2 \, t}}{(\sigma \, \sqrt{2 \pi t})^d}\, e^{-\frac{|x-B\,t|^2}{2\, \sigma^2 \, t}}, \hbox{ for }t>0.
\end{equation}
Regarding the observations, we focused on the dispersing particles, and considered $3$ observation times $t_1=1$, $t_2=3$, $t_3=6$ and $3$ observation windows $A_1=(0,1)\times (0,1)$, $A_2=(-1,0)\times (-1,0)$ and $A_3=(1,2)\times(2,3)$, see Fig.~\ref{fig:particles}.

The unknown parameters are $\Theta=(B(1),B(2), \sigma,\kappa)$. The parameters are estimated by maximising the pseudo-likelihood $\pLl(\Theta)$ in \eqref{eq:like1} when using the full spatial observation data, and by using  $\pLl_c(\Theta)$ in \eqref{eq:like_count} when the information is restricted to the counting. We assumed the following constraints on the parameter values: $\Theta\in \Gamma:=(-1,1)\times (-1,1) \times (0.01,1) \times (0.1,10).$ We used the Matlab$^\copyright$ built-in gradient-based minimization algorithm fmincon$^\copyright$ with default options, and applied to $-\log(\pLl)$ to compute the maximum pseudo-likelihood estimators (MPLE). 
The results are summarized in Table~\ref{tab:MLE_space}. As expected, the observation of the positions of the particles leads to a more accurate estimation compared to counting data. In addition, in this example, the use of the pseudo-likelihood (instead of the likelihood) and the simplifying approximation concerning the initial distribution of particles provides a quite satisfactory estimation accuracy when particle locations are observed.

\begin{table}
\center
\begin{tabular}{cccc}
 \hline	\hline
   Parameter & True value & MPLE - locations & MPLE - counting \\
 \hline
   $B(1)$ & 0.3 & 0.30 (0.02)& 0.36 (0.10)\\
 $B(2)$ & 0.5 & 0.50 (0.02)  & 0.56 (0.07)\\
 $\sigma$ & 0.3 & 0.30 (0.02)  & 0.27 (0.03) \\
 $1/\kappa^2$ & 3 & 3.04 (0.23) & 8.34 (2.28) \\
  \hline \hline
 \end{tabular}
  \caption{Mean and standard deviation (based on $10^3$ repetitions) of the maximum pseudo-likelihood estimators (MPLE) obtained from observations consisting of either particle locations (using $\pLl(\Theta)$ in \eqref{eq:like1}) or counting data (using $\pLl_c(\Theta)$ in \eqref{eq:like_count}).  }\label{tab:MLE_space}
\end{table}

\section{Discussion \label{sec:disc}}

In this work, we have revisited the SPDE approach now commonly used in spatial statistics, as proposed in \citet{LinRue11} and further developed in \citet{CarAll21}. We have adopted a mechanistic perspective based on the movement of microscopic particles, which led us to relate pseudo-differential operators to  dispersal kernels. This mechanistic approach led us to a new and deeper understanding of a certain class of models and paved the way to new models with original features that do not fit in the general framework of \citet{CarAll21} based on Fourier transforms, such as drifts and nonlinear reaction terms. We also showed on a synthetic case study how such mechanistic models, associated to a probabilistic observation model, can be used in a hierarchical setting to estimate the parameters of the particle dynamics. This approach has a large potential in ecology and epidemiology for the studies of the dynamics of populations that can be described at the resolution of individuals \citep[animals, seeds, pollen grains, pathogens, spores;][]{tufto1997,klein2003,dragon2012,soubeyrand2015}.

Regarding stationary models on which a Fourier analysis is possible,  we provided in Section \ref{sec:dispersal} a detailed exposition of several dispersal operators, with the aim to bridge the gap between the mechanistic view (through dispersal kernels) and the statistical view (through covariance functions). Our main findings are summarised in Table \ref{tab:ccl}. We first have shown that the solution of the evolving Mat\'ern equation~\eqref{eq:evol_Matern} is driven by three main forces: dispersal through a thin-tailed dispersal kernel; absorption through a negative linear reaction term; and additive noise. The spatial symbol function is $g(\xi) = (\|\xi\|^2 + \kappa^2)^{\alpha/2}$ and the associated covariance function is a Matérn function which decays exponentially fast as $\|h\|\to \infty$. We then turned our attention to the much less usual fractional Laplacian operator with a linear reaction term. In this case,  the dispersal is driven by a fat-tailed kernel, and the corresponding symbol function is $g(\xi) = \|\xi\|^{\alpha}+ \kappa^2$. The associated covariance functions describe long-distance dependencies,  see Eqs \eqref{eq:first_aux} and \eqref{eq:rho_struve_final}. The difference between these covariance functions and the usual Matérn covariance is illustrated in Fig.~\ref{fig:covar}. The difference between the thin-tailed and fat-tailed long distance behaviours is clearly visible, even though both covariances are comparable at short positive distances. Notice that the two operators are equal when $\alpha=2$, which corresponds to the usual Laplacian operator with linear reaction. Finally, we also considered integral dispersal operators which are directly defined trough a convolution with a dispersal kernel. This makes it possible to consider very general dispersal tails, from very thin (e.g., Gaussian) to very fat (e.g., with algebraic decay). With such operators, the computations of the symbol function $g$ and therefore of the spectral measure are straightforward. We derived explicit (spatial) covariance functions for exponential kernels which, as expected, are closely related to the Matérn family.  When the kernel is a general Matérn function (with integer parameter $\nu$), it is shown that the associated Gaussian random field is the sum of Markov random fields.  As illustrated with the fractional Laplacian operator, other types of kernels should lead to other families of covariance functions. However, the derivation of explicit expressions is not straightforward in general.

When a Fourier analysis is not possible, for instance with spatially heterogeneous coefficients or with a nonlinear reaction term, the framework of \cite{CarAll21} cannot be applied as such. Nevertheless, the mechanistic approach that we have proposed allows to construct spatio-temporal random fields, based on an intuitive approach. Of course, their mathematical characterisation should rely on much more work and is far beyond the scope of our work. Still, the examples in Sections~\ref{sec:drift} and~\ref{sec:bistable} illustrate the interest of this intuitive approach. First, to construct spatio-temporal random fields which reflect the effect of external driving forces, such as the effect of wind on a concentration.  It appears in the simulations of Section~\ref{sec:drift} that adding a spatio-temporally varying drift term in the equations leads to a particular form of spatio-temporal patterns which are consistent with the underlying physical assumptions. When such external constraints are known, it seems essential to take them into account for the realism of the model. Second, to tackle the case with several attracting states. For instance, as we explained in Section~\ref{sec:bistable}, many biological systems have bistable dynamics, with typically the steady state $0$ (the absence of the species) and another positive steady state. We proposed here to construct spatio-temporal random fields which reflect these dynamics through stochastic Allen-Cahn equations. These equations must be used with care, as their well-posedness depends on the dimension and on the noise term. Nevertheless, our preliminary computations show that the characteristics of the corresponding random fields meet our expectations, especially regarding the bimodality of the distributions. This advocates for further mathematical studies of these equations.

In Section~\ref{sec:drift}, we have considered spatio-temporally varying drift terms. Spatio-temporally varying diffusion coefficients may be considered as well. Two main types of diffusion operators are found in the reaction-diffusion literature: the Fokker-Planck diffusion operator $\D(t,x,[u])=\Delta(D(x,t)u)$,  and the Fickian diffusion operator $\D(t,x,[u])=\hbox{div}(D(x,t)\nabla u)$ \citep{Roq13}, see also \cite{AlfGil21} for generalisations of the form $\D(t,x,[u])=\hbox{div}(D(x,t)^q\nabla( D(x,t)^{1-q} u))$ ($q\in(0,1)$). Both of them have anisotropic versions (see Remark~\ref{rem:aniso}), but for simplicity we consider here the isotropic case. We have observed in Section~\ref{sec:Levy} that the Fokker-Planck diffusion emerges as the macroscopic limit of a stochastic diffusion process with spatially-dependent diffusion term $\sigma(x,t)=\sqrt{2 \, D(x,t)}$. On the other hand Fickian diffusion is usually derived from physical relationships between particle flux, concentration and concentration gradient,   in which the driving force along time is that a system seeks equilibrium. If in some elementary space-time volume the density is higher than in the neighbouring area, particles will move from the denser area to the less dense ones.  More precisely, the flux is assumed to be linearly linked to the particle concentration gradient (Fick's Law), with a coefficient $D(x,t)$ that accounts for the spatial variations of the environment. Thus, the Fokker-Planck diffusion operator is generally better suited to the description of particle movements \citep[see e.g.,][]{HanEkb01,RoqAug08}, while the Fickian diffusion operator is more suitable for the description of physical phenomena, such as the dilution of a dye in a liquid or heat diffusion in a heterogeneous body \cite{Fic55}. 
We can note that Fokker-Planck diffusion equation can also be written:
$\Delta (D \, u) = \hbox {div} (\nabla [D \, u]) = \hbox {div} (D \, \nabla u) + \hbox {div} (u \nabla D).$ We thus obtain a Fickian diffusion with a drift term oriented opposite the gradient of $D$. Thus, these two operators, which are identical when the coefficient $ D (t, x) $ is constant, lead to different behaviors of the solution  in a heterogeneous environment. Fickian tends to homogenize the solution, while with the Fokker-Planck equation, the solution will tend to concentrate in the regions where $ D $ is small, due to the drift  term $ \hbox {div} (u \nabla D). $ This is consistent with the above explanations: for particles, one can expect to find a greater concentration in regions of low mobility; conversely, for heat diffusion, a homogenization phenomenon is expected, whatever the diffusion in the media considered. We briefly checked numerically whether some obvious differences could be observed on the solutions of \eqref{eq:SPDE_gale}, 
but did not observe significant differences,
see Appendix~B. This finding is in strong contrast with the deterministic case, where the solutions tend at large times to become proportional to $1/D(x)$ (if $D$ does not depend on time) with a Fokker-Planck diffusion vs to a constant with a Fickian diffusion.

In our opinion, this preliminary work opens new avenues for research in an effort to further tie together spatial statistics and PDEs. We see at least three main lines of research. First, as already pointed out above, more effort should be made to propose new SPDEs with physical or biological interpretations: including drifts, non homogeneous parameters and nonlinear reactions are key for a flexible and useful modelling. However, the theoretical mathematical behaviour is unclear or unknown in some cases, but numerical solutions are possible. In this work, the exponent $\alpha$ was set to be less than or equal to 2. While the Fourier definition $\F(-(-\Delta)^{\alpha/2}u)(\xi)=-\|\xi\|^\alpha \F(u)(\xi)$ allows to consider values $\alpha>2$, the mechanistic construction of the fractional Laplacian proposed in Section~\ref{sec:Levy} and the convolution formula \eqref{eq:pointwise_frac_laplace} cannot be used for such values of $\alpha$. Another approach uses hypersingular integrals based on higher order finite differences (at least of order $\lceil \alpha \rceil$) instead of the first order finite difference in \eqref{eq:pointwise_frac_laplace} \citep[see Part 1, Chapter 3 in][]{Sam01}. However, the interpretation of this operator in terms of particle motion is far from obvious. 

As a second line of research, algorithmic developments are required for these models, in particular in a spatio-temporal context where datasets can be large or massive. Using the Garlekin method as in \citet{LinRue11} is one possible option, but care must be taken because the Markov property can be lost for the most complex models. The numerical analysis community has developed tools for solving PDEs that could be of great use to the statistical community, in particular for spatial prediction and stochastic simulations.

As a third line of research, we advocate the use of hybrid models in the spirit of the application shown in Section \ref{sec:PDEs} combining (presumably in a hierachical setting) mechanistic, probabilistic and statistical compartments, on which inference of the parameters can be made in a frequentist or in a Bayesian context.
Interestingly, the proposed approach leads to spatial and spatio-temporal point processes with a mechanistic foundation and an interpretable intensity function. These point processes, as ordinary point processes, can be fitted to point patterns, counting data, and even information about the presence or absence of points. However, due to their construction, we can also use trajectory data to estimate their parameters. Whatever the sampling density of the trajectories, the crucial feature is the ability to follow certain points in time, that is to say to have (at least partial) information about the $\psi_k$s in Eqs \eqref{eq:XgivenXm1}--\eqref{eq:XKgivenXKm1}. Such data are expected to bring accurate information about the parameters but, for fully exploiting their information content, the true likelihood \eqref{eq:likelihood} should be used, or at least a new pseudo-likelihood,  yet to be proposed,  accounting for full temporal dependencies in the locations of particles that are tracked across time. It would also be interesting to combine several types of data (locations, counting, trajectories) and even being able to advise how to mix them for maximising the information they bring for a limited total sampling effort. This task could be carried out with the study of the Fisher information, at least in simple cases where the expression of $u$ is explicit like in Equation \eqref{eq:expr_u_homo4}. Finally, including direct dependencies between trajectories of particles (e.g., attraction of  particles as in \citet{soubeyrand2011} or repulsion) deserves to be studied to gain in modelling realism and to be able to handle more complex dynamics. In this case, the challenges would be to derive (i) the mechanistic equations satisfied by the intensity function \citep[this could take a similar course to the Keller-Segel approach, possibly with fractional diffusion as in][]{BouCal10}, (ii) the nature of the resulting point processes, which are expected to be more complex than those obtained in the case of independent trajectories, and (iii) an estimation approach that would allow us to achieve accurate parameter estimators.

\bibliographystyle{chicago}
\bibliography{biblio_lionel,biblio_denis,biblio_samuel}

\newpage

\appendix

\section{Point process generated by the SDE-based model of dispersing particles}

\subsection{Reminder of and supplementary notations.}

$\X_t$ is the process of locations of still dispersing particles at time $t\ge 0$, $N_0$ the initial number of particles that is Poisson-distributed with mean $\Nzb$, and $\tau_i$ the deposition time of particle $i$ ($\tau_i$s are independently and identically distributed from the exponential distribution with mean $1/\kappa^2$). The initial locations (given by $\mathbf{X}_0$) are independently and identically distributed from the probability distribution function $p_0$, which can be either discrete or continuous on the space $\R^d$. When a particle has been deposited, it is no more included in the point process $\mathbf{X}_t$.

$A,A_1,\ldots,A_K$ are non-empty Lebesgue-measurable subsets of $\R^d$, $K\in\N^*$, $0<s<t$ and $0<t_1<\ldots<t_K$. $A^c$ is the complementary set of $A$, i.e., $A\cup A^c=\R^d$ and $A\cap A^c=\emptyset$. For $t\in[0,\tau_i)$, $X_t^i$ is the location of particle $i$ at time $t$. $\X_{tA}$ is the restriction of $\mathbf{X}_t$ over $A$. $I_{tA}$ is the subset of indices in $\{1,\ldots,N_0\}$ corresponding to the points $\X_{tA}$. $N_t(A)$ is the number of particle located in $A$ at $t$. We use the notation $\pr$ to denote the probability measure of variables or sets of variables. Conditional probability measures are written $\pr(\cdot\mid\cdot)$. We use the abbreviation p.d.f. for probability distribution function.

For any $s\ge 0$ and any distribution $q$ on $\R^d$, we denote by $u^s_q(\cdot,t)$ the solution of Eq.~\eqref{eq:FP_Levy_u} starting at time $s$ with the condition $u(\cdot,s)=q(\cdot)$. Similarly, we denote by  $p^s_q(\cdot,t)$ the solution of Eq.~\eqref{eq:FP_Levy} starting at time $s$ with the condition $p(\cdot,s)=q(\cdot)$. We note that the two quantities are related by the relationship $u^s_q(\cdot,t)=e^{- (t-s)\kappa^2}p^s_q(\cdot,t)$ for all $t\ge s$. When $q$ is the sum of Dirac measures centred at the locations of the particles $\X_{sA}$,  $q(x)=\sum_{i=1}^{N_s(A)}\delta(\cdot-X_{sA}^i)$, we simply write $u^s_{\X_{sA}}(\cdot,t)$ (resp. $p^s_{\X_{sA}}(\cdot,t)$).%For simplicity and consistency with the other parts of the manuscript, we denote by $u(\cdot,t)$ the solution of Eq.~\eqref{eq:FP_Levy_u} starting at $0$ with the initial condition $\Nzb p_0$; namely $u(\cdot,t)=u^0_{\Nzb p_0}(\cdot,t)$.

\subsection{Lemmas}

\begin{lemma}\label{lem:poisson}
For any fixed time $t>0$, the spatial point process $\X_t$ is an inhomogeneous Poisson point process with intensity $u^0_{\Nzb p_0}(\cdot,t)$.
\end{lemma}

$\X_t$ is a spatial Poisson point process with intensity function $\rho$ defined over $\R^d$ if for any bounded region $A \subset \R^d$, $N_t(A)$ is Poisson distributed with mean $\int_A \rho(x)dx$, and conditioned on $N_t(A) = n$, the 
$n$ events in $A$ are independent and identically distributed with density proportional to $\rho$ restricted to $A$. 
The elements of proof for Lemma \ref{lem:poisson} are provided in the main text (Section~\ref{sec:PDEs}).

\begin{lemma}\label{lem:dependence}
For any fixed times $t>s>0$, the spatial point process $\X_t$ conditional on the partial observation $\X_{sA}$ of $\mathbf{X}_s$ restricted to $A$, is not an inhomogeneous Poisson point process.
\end{lemma}

\noindent {\it Proof of Lemma \ref{lem:dependence}.} 
Let $A$ be an open bounded set and assume that $N_s(A)>0$. Let $B\subset A$ an open set such that $N_s(B)=1$, and denote  $X_t^1 \in \R^d$ the position of this single particle at times $t\ge s$. According to Section~\ref{sec:Levy}, the expected particle density at times $t\ge s$ is $u^s_q(\cdot,t)$, with the initial condition (here, at time  $s$), $$q(x) =\delta(x-X_s^1)+\tilde{u}(x), \hbox{ with }\ds \int_{B}\tilde{u}(y) \, dy=0,$$ where $\delta$ is a Dirac measure at $0$. By continuity of $t\mapsto \int_{B}u(y,t) \, dy$ at $t=s$, this implies that $ \lim\limits_{t\to s^+}E[N_t(B)]=\int_{B}u(y,s) \, dy=1$. Assume that the spatial point process $\X_t$ conditional on the partial observation $\X_{sA}$ of $\mathbf{X}_s$ restricted to $A$ is  an inhomogeneous Poisson point process for $t>s$. Then, as $\lim\limits_{t\to s^+}E[N_t(B)] =1$, we have 
\begin{equation}
    \label{eq:lim_poisson}
    \lim\limits_{t\to s^+} P(N_t(B)\ge 2)=1-2 \, e^{-1}.
\end{equation}
On the other hand, the probability $P(X_t^1\in B \hbox{ and }t<\tau_1)$ that $X_t^1$ is still dispersing and belongs to $B$ at time $t\ge s$ is given by $\int_B h(y,s) \, dy$, with $h$ the solution of Eq.~\eqref{eq:FP_Levy_u}, this time with initial condition $h(x,s)=\delta(x-X_s^1)$. By continuity, one has $\lim\limits_{t\to s^+}P(X_t^1\in B \hbox{ and }t<\tau_1)=1$. Writing
$$E[N_t(B)]\ge P(X_t^1\in B \hbox{ and }t<\tau_1) + P(\exists \ i\neq 1 \hbox{ s.t. }X_t^i\in B \hbox{ and }t<\tau_i),$$and passing to the limit $t\to s^+$, we obtain  that $\lim\limits_{t\to s^+}  P(\exists \ i\neq 1 \hbox{ s.t. }X_t^i\in B \hbox{ and }t<\tau_i)=0$. This contradicts \eqref{eq:lim_poisson}.
$\Box$

Intuitively, the non-Poisson nature of $\X_t\mid\X_{sA}$ can be understood by looking at the variance of $N_{s^+}(A)$  when $t=s^+$ ($0<s^+-s \ll 1$). Assume that $N_{s}(A)\ge 1$. We have  $V(N_{s^+}(A)\mid\X_{sA})\approx V(N_s(A)\mid\X_{sA})= 0$. On the other hand, if $\X_t\mid\X_{sA}$ was a Poisson process, we would have $V(N_{s^+}(A)\mid\X_{sA})=E(N_{s^+}(A)\mid\X_{sA})\approx N_{s}(A)$. Hence, given $X_{sA}$, the number of points in $A$ at $s^+$ cannot be Poisson-distributed.

%Intuitively, the non-Poisson nature of $\X_t\mid\X_{sA}$ can be understood by looking at the counts of $X_t$--points belonging to $A$ and $A^c$ when $t=s^+$ ($0<s^+-s<<1$). At time $s^+$, $N_{s^+}(A)\approx N_s(A)$ and $N_{s^+}(A^c)\approx N_s(A^c)$, $E(N_{s^+}(A)\mid\X_{sA})\approx E(N_s(A)\mid\X_{sA})=N_s(A)$ and $V(N_{s^+}(A)\mid\X_{sA})\approx V(N_s(A)\mid\X_{sA})= 0$, whereas $E(N_{s^+}(A^c)\mid\X_{sA})\approx E(N_s(A^c)\mid\X_{sA})=E(N_s(A^c)) \Li =\int_{A^c} u^0_{\Nzb p_0}(x,s) \, dx$\Bk{}  and $V(N_{s^+}(A^c)\mid\X_{sA})\approx V(N_{s}(A^c)\mid\X_{sA})=V(N_{s}(A^c))=\Nzb\int_{A^c}u^0_{\Nzb p_0}(x,s) \, dx$\Bk{}. Hence, given $X_{sA}$, the number of points in $A$ at $s^+$ cannot be Poisson-distributed.

\begin{lemma}\label{lem:union}
For any fixed times $t>s\ge 0$, the point process $\X_t$ conditional on the partial observation $\X_{sA}$ of $\mathbf{X}_s$ restricted to $A$, is the union of a Poisson point process with intensity  $u_{u^0_{\Nzb p_0}(\cdot,s)\1_{A^c}(\cdot)}^s(\cdot,t)$  and 
$N_s(A)$ independent thinned binomial point processes with size 1, thinning probability $1-\exp(-(t-s)\kappa^2)$, and p.d.f. $p_{\delta(\cdot-X_{sA}^i)}^s(\cdot,t)$, 
where
$X_{sA}^i$ is the point of $\X_{sA}$ with label $i$ and $\1_v$, $v\subset \R^d$, is the indicator function of $v$.
\end{lemma}

\noindent {\it Proof of Lemma \ref{lem:union}.} Each point in $\X_{sA}$ independently generates at time $t$ either zero point or one point. It generates zero point if the particle is deposited between $s$ and $t$, which occurs with probability $1-\exp(-(t-s)\kappa^2)$. Otherwise, it generates one point whose probability distribution function over $\R^d$ is  $p_{\delta(\cdot-X_{sA}^i)}^s(\cdot,t)$. %The integral of this intensity decreases with time (after $s$) because of the non-conservative term $- \kappa^2 u$ in Eq.~\eqref{eq:FP_Levy_u} which explains the probability $1-\exp(-(t-s)\kappa^2)$ that $X_{sA}^i$ may lead to zero point at time $t$. Hence, if the particle at $X_{sA}^i$ at time $s$ is still dispersing at time $t$, its p.d.f. is $u_{X_{sA}^i}^s(\cdot,t)/\int_{\R^d} u_{X_{sA}^i}^s(\cdot,t){\color{red}=u_{X_{sA}^i}^s(\cdot,t)/\exp(-(t-s)\kappa^2)}$.

From Lemma \ref{lem:poisson}, $\X_s$ is an inhomogeneous Poisson point process with intensity $u^0_{\Nzb p_0}(\cdot,s)$.
Setting $\X_s=\X_{sA}+\X_{sA^c}$ and because of the Poisson nature of $\X_s$, $\X_{sA}$ and $\X_{sA^c}$ are independent and $(\X_{sA^c}\mid\X_{sA})\equiv \X_{sA^c}$ in distribution. Hence, $\X_{sA^c}\mid\X_{sA}$ is an inhomogeneous Poisson point process with intensity
$u^0_{\Nzb p_0}(\cdot,s)\1_{A^c}(\cdot)$. It follows that the number of dispersing particles in $A^c$ at $s$ is drawn from a Poisson distribution with mean $\int_{A^c} u^0_{\Nzb p_0}(\cdot,s)$, and the particle depositions are i.i.d. from the p.d.f. $u^0_{\Nzb p_0}(\cdot,s)\1_{A^c}(\cdot)/\int_{A^c} u^0_{\Nzb p_0}(\cdot,s)$. Hence, we can apply Lemma \ref{lem:poisson} between times $s$ and $t$ to determine the distribution of the part of $\X_t$ that was restricted to $A^c$ at time $s$. This part of $\X_t$ is an inhomogeneous Poisson point process with intensity $u_{E(\X_{sA^c})}^s(\cdot,t)= u_{u^0_{\Nzb p_0}(\cdot,s)\1_{A^c}(\cdot)}^s(\cdot,t).$ 
$\Box$

\begin{corollary}\label{cor:intensity}
For any fixed time $t>0$, the intensity function of the point process $\X_t$ conditional on the partial observation $\X_{sA}$ of $\mathbf{X}_s$ restricted to $A$ is $x\mapsto u_{\X_{sA}}^s(x,t)+ u_{u^0_{\Nzb p_0}(\cdot,s)\1_{A^c}(\cdot)}^s(x,t)$ whose expectation is $u^0_{\Nzb p_0}(\cdot,t)$.
\end{corollary}

\noindent {\it Proof of Corollary \ref{cor:intensity}.} 
The first part of the proof of Lemma \ref{lem:union} yields that the sub-point process $\X_{sA}$ generates at $t$ a (non-Poisson) point process with conditional intensity
$\sum_{i=1}^{N_s(A)}  u_{\delta(\cdot-X_{sA}^i)}^s(x,t)= u_{\X_{sA}}^s(x,t)$
given $\X_{sA}$. Moreover, from Lemma \ref{lem:union}, the remaining points in $\X_t$ (i.e., the locations at $t$ of particles that were in $A^c$ at $s$) have intensity $u_{u^0_{\Nzb p_0}(\cdot,s)\1_{A^c}(\cdot)}^s(x,t)$. The sum of the two intensities gives the intensity of $\X_t$ given $\X_{sA}$. This is clear with the following reasoning: Let $\Lambda$ denote the intensity of $\X_t$ conditional on $\X_s$: $\Lambda(x)=u_{\X_s}^s(x,t)=u_{\X_{sA}}^s(x,t)+u_{\X_{sA^c}}^s(x,t)$. The expectation of $\Lambda$ given $\X_{sA}$ is the intensity of $\X_t\mid\X_{sA}$ and satisfies, because of the linearity of Eq.~\eqref{eq:FP_Levy_u}:
\begin{align*}
E(\Lambda(x)\mid \X_{sA})
    &=u_{\X_{sA}}^s(x,t)+E(u_{\X_{sA^c}}^s(x,t))\\
    &=u_{\X_{sA}}^s(x,t)+u_{E(\X_{sA^c})}^s(x,t) \\
    &=u_{\X_{sA}}^s(x,t)+ u_{u^0_{\Nzb p_0}(\cdot,s)\1_{A^c}(\cdot)}^s(x,t) \\
    &=u_{\sum_{i=1}^{N_s(A)}\delta(\cdot-X_{sA}^i)+u^0_{\Nzb p_0}(\cdot,s)\1_{A^c}(\cdot)}^s(x,t) \\
    &= u^0_{\Nzb p_0}(x,t) + 
\Big(u_{\X_{sA}}^s(x,t)-u^s_{ u^0_{\Nzb p_0}(\cdot,s)\1_{A}(\cdot)}(x,t)\Big).
\end{align*}
The third line above gives the expression of the intensity mentioned in the corollary. From the fifth line and the linearity of Eq.~\eqref{eq:FP_Levy_u} we get:
\begin{align*}
E(\Lambda(x))
    &=u^0_{\Nzb p_0}(x,t) + 
\Big(E(u_{\X_{sA}}^s(x,t))-u^s_{ u^0_{\Nzb p_0}(\cdot,s)\1_{A}(\cdot)}(x,t)\Big)\\
    &=u^0_{\Nzb p_0}(x,t) + 
\Big(u_{E(\X_{sA})}^s(x,t)-u^s_{ u^0_{\Nzb p_0}(\cdot,s)\1_{A}(\cdot)}(x,t)\Big)\\
    &= u^0_{\Nzb p_0}(x,t),
\end{align*}
which is consistent with Lemma \ref{lem:poisson} that states that $ u^0_{\Nzb p_0}(\cdot,t) $ is the unconditional intensity of $\X_t$.
$\Box$

\subsection{Likelihood function}
Our goal is to prove the formula~\eqref{eq:likelihood}. We begin by writing:
\begin{align*}
\Ll(\Theta)&=\int_{\tilde\X^{(K-1)}} d\pr(\X^{(1)},\ldots,\X^{(K)},\tilde\X^{(K-1)})\\
&= \int_{\tilde\X^{(K-1)}} 
\pr(\X^{(K)}\mid \X^{(1)},\ldots,\X^{(K-1)},\tilde\X^{(K-1)})
d\pr(\X^{(1)},\ldots,\X^{(K-1)},\tilde\X^{(K-1)})\\
&= \int_{\tilde\X^{(K-1)}} 
\pr(\X^{(K)}\mid \X^{(1)},\ldots,\X^{(K-2)},\X_{t_{K-1}})
d\pr(\X^{(1)},\ldots,\X^{(K-2)},\X_{t_{K-1}})\\
&= \int_{\tilde\X^{(K-1)}} 
\pr(\X^{(K)}\mid \X_{t_{K-1}})
d\pr(\X^{(1)},\ldots,\X^{(K-2)},\X_{t_{K-1}}).
\end{align*}
Then, consider the property $(\mathcal{P}_j)$:
$$\Ll(\Theta)= \int_{\tilde\X^{(K-1)},\ldots,\tilde\X^{(K-1-j)}}
\pr(\X^{(K)}\mid \X_{t_{K-1}})
\left(\prod_{k=K-j}^{K-1}d\pr(\X_{t_{k}}\mid\X_{t_{k-1}})\right)
d\pr(\X^{(1)},\ldots,\X^{(K-2-j)},\X_{t_{K-1-j}}).
$$The above computations show that the property $(\mathcal{P}_0)$ is true (with the convention that the product equals 1 if $j=0$).
%(with the convention that the product over an empty set is equal to $1$).
Assume that $(\mathcal{P}_j)$ is satisfied for some integer $j \in \{0,\ldots, K-4\}$. Then
\begin{multline*}
    \Ll(\Theta) = \int_{\tilde\X^{(K-1)},\ldots,\tilde\X^{(K-2-j)}}
\pr(\X^{(K)}\mid \X_{t_{K-1}})
 \left(\prod_{k=K-j}^{K-1}d\pr(\X_{t_{k}}\mid\X_{t_{k-1}})\right) \\ 
d\pr(\X^{(1)},\ldots,\X^{(K-2-j)},\tilde\X^{(K-2-j)},\X_{t_{K-1-j}}), 
\end{multline*}
\begin{multline*}
 =\int_{\tilde\X^{(K-1)},\ldots,\tilde\X^{(K-2-j)}}
\pr(\X^{(K)}\mid \X_{t_{K-1}})
 \left(\prod_{k=K-j}^{K-1}d\pr(\X_{t_{k}}\mid\X_{t_{k-1}})\right) \\
d\pr(\X^{(1)},\ldots,\X^{(K-3-j)},\X_{t_{K-2-j}},\X_{t_{K-1-j}}), 
\end{multline*}
\vspace{-10mm}
\begin{multline*}
 =\int_{\tilde\X^{(K-1)},\ldots,\tilde\X^{(K-2-j)}}
\pr(\X^{(K)}\mid \X_{t_{K-1}})
 \left(\prod_{k=K-j}^{K-1}d\pr(\X_{t_{k}}\mid\X_{t_{k-1}})\right) \\
d\pr(\X_{t_{K-1-j}}\mid \X^{(1)},\ldots,\X^{(K-3-j)},\X_{t_{K-2-j}}) 
 d\pr( \X^{(1)},\ldots,\X^{(K-3-j)},\X_{t_{K-2-j}}) 
\end{multline*}
\vspace{-10mm}
\begin{multline*} =\int_{\tilde\X^{(K-1)},\ldots,\tilde\X^{(K-2-j)}}
\pr(\X^{(K)}\mid \X_{t_{K-1}})
 \left(\prod_{k=K-j-1}^{K-1}d\pr(\X_{t_{k}}\mid\X_{t_{k-1}})\right) \\
 d\pr( \X^{(1)},\ldots,\X^{(K-3-j)},\X_{t_{K-2-j}}).
\end{multline*}
Thus, property $(\mathcal{P}_{j+1})$ is true. By induction, this shows that $(\mathcal{P}_{K-3})$ is true, i.e.:
\begin{align*}
\Ll(\Theta)&= \int_{\tilde\X^{(K-1)},\ldots,\tilde\X^{(2)}}
\pr(\X^{(K)}\mid \X_{t_{K-1}})
\left(\prod_{k=3}^{K-1}d\pr(\X_{t_{k}}\mid\X_{t_{k-1}})\right)
d\pr(\X^{(1)},\X_{t_2}).
\end{align*}
Finally, integrating the right-hand-side of $(\mathcal{P}_{K-3})$ with respect to $\tilde\X^{(1)}$ leads to:
\begin{align*}
\Ll(\Theta)&= \int_{\tilde\X^{(K-1)},\ldots,\tilde\X^{(1)}}
\pr(\X^{(K)}\mid \X_{t_{K-1}})
\left(\prod_{k=3}^{K-1}d\pr(\X_{t_{k}}\mid\X_{t_{k-1}})\right)
d\pr(\X^{(1)},\tilde\X^{(1)},\X_{t_2})\\
&= \int_{\tilde\X^{(K-1)},\ldots,\tilde\X^{(1)}}
\pr(\X^{(K)}\mid \X_{t_{K-1}})
\left(\prod_{k=3}^{K-1}d\pr(\X_{t_{k}}\mid\X_{t_{k-1}})\right)
d\pr(\X_{t_1},\X_{t_2})
\end{align*}
and, by writing $d\pr(\X_{t_1},\X_{t_2})=d\pr(\X_{t_2}\mid\X_{t_1})d\pr(\X_{t_1})$, we get formula~\eqref{eq:likelihood}.

\section{Fokker-Planck and Fick's law}
\begin{figure}[h!]
\center
\subfigure[Fokker-Planck]{\includegraphics[width=0.45\textwidth]{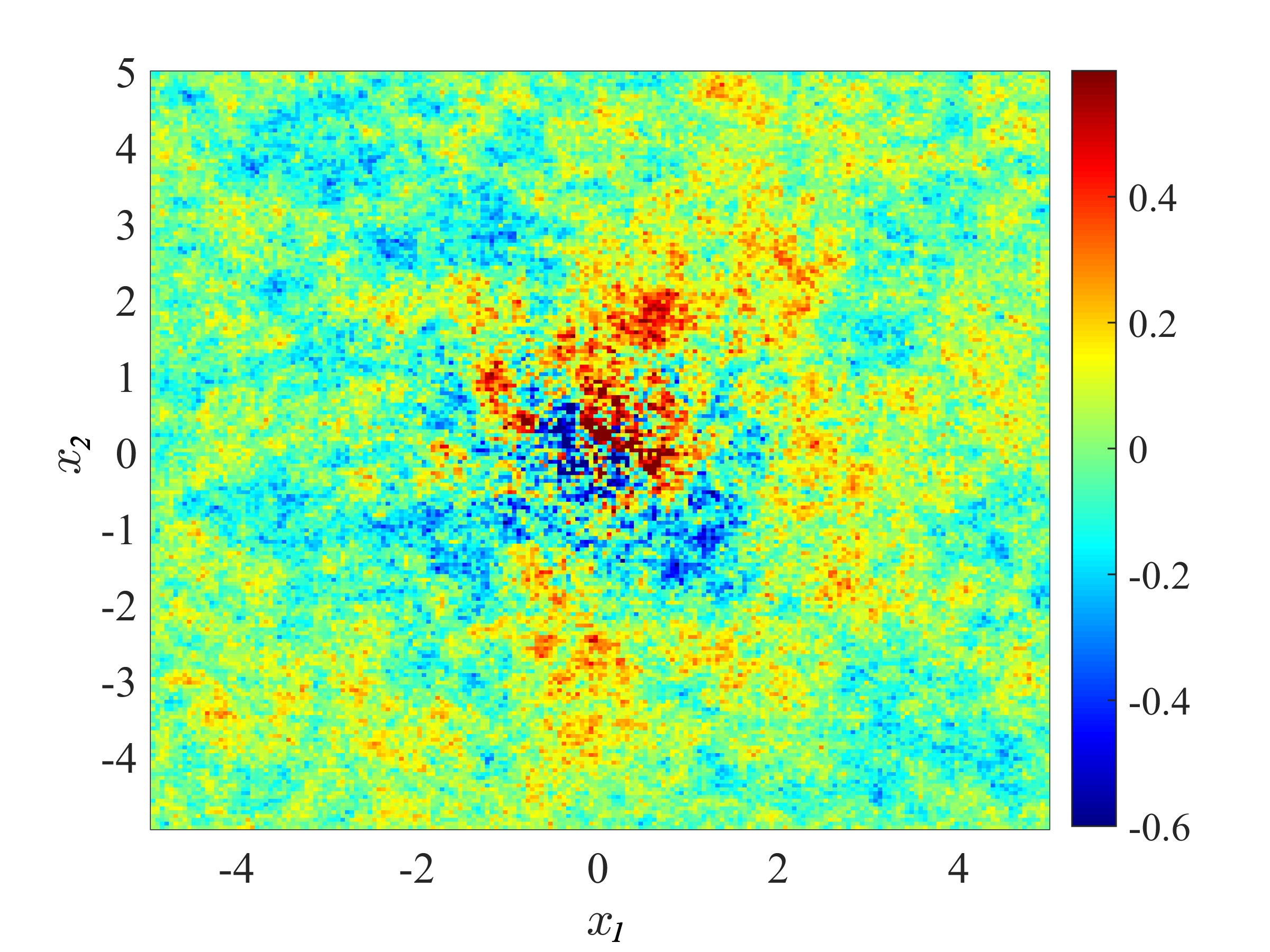}}
\subfigure[Fick]{\includegraphics[width=0.45\textwidth]{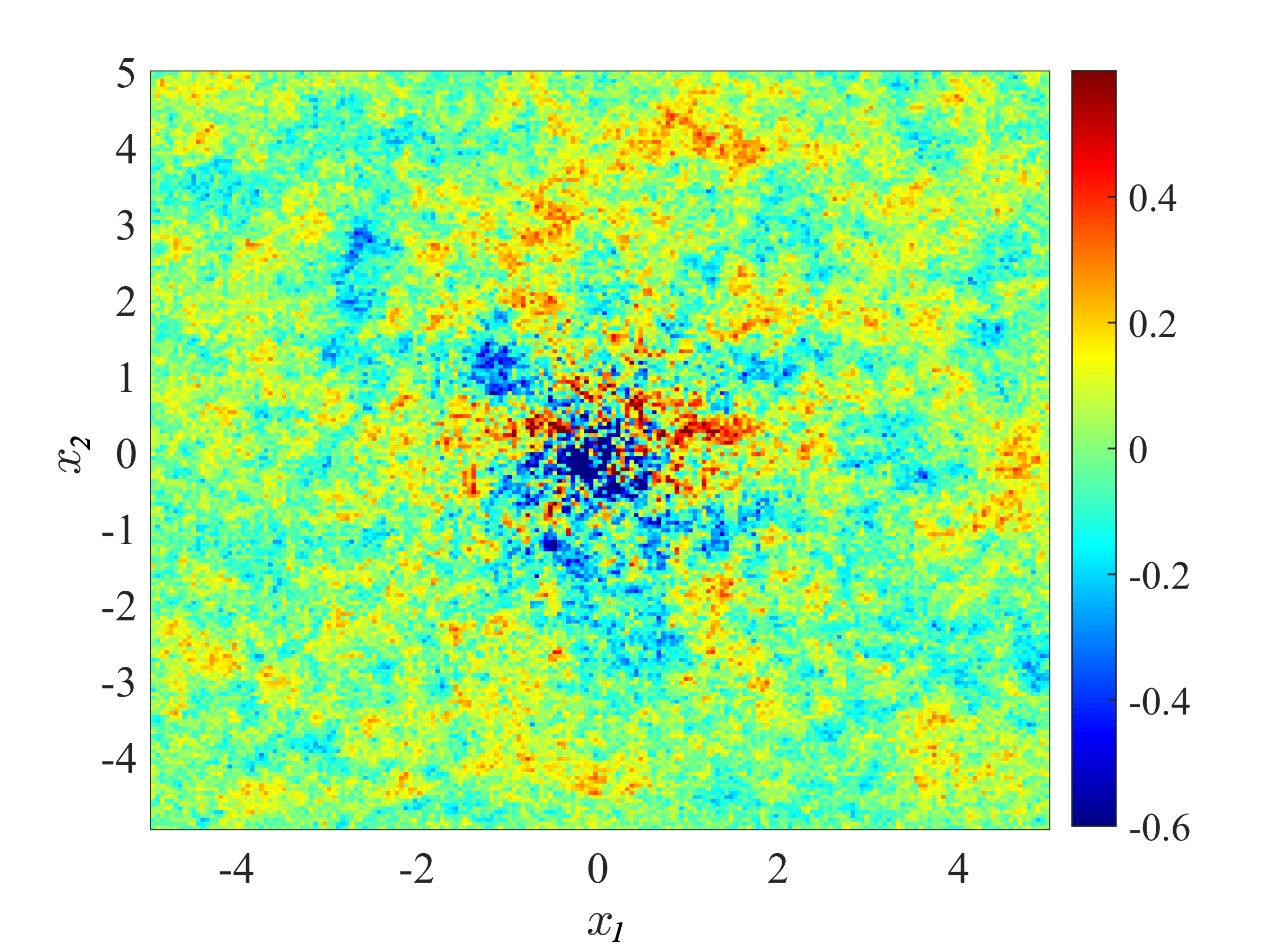}}
\caption{{\bf Fokker-Planck vs Fickian diffusion.} Solution of the SPDE \eqref{eq:SPDE_gale} with dispersal term $\D(t,x,[u])=\Delta(D(x,t)u)$ (panel a) vs $\D(t,x,[u])=\hbox{div}(D(x,t)\nabla u)$ (panel b), reaction term $f(t,x,u)=-\kappa^2 \, u(x,t)$ (absorption) and drift term $\B (t,x,[u])=0$. The diffusion coefficient is $D(x,t)=D(x)=D_0+D_1(1-e^{-\|x\|^2/(2 \sigma_D^2)})$, with $D_0=10^{-3}$ and $D_1=10^{-1}.$ In both cases $\kappa=0.1$, $\sigma=1$ and the solution is plotted at a fixed time $T=10$.}
\label{fig:FPvsFick}
\end{figure}

We computed the solution of \eqref{eq:SPDE_gale}
with a null drift term $\B (t,x,[u])=0$ and an absorption term $f(t,x,u)=-\kappa^2 \, u(x,t)$, depending on the type of dispersal operator $\D$ of either Fokker-Planck or Fickian form. The results in Fig.~\ref{fig:FPvsFick}, and other preliminary computations (Matlab codes: \url{https://osf.io/w5utd/}) did not show significant differences between the solutions obtained with the two operators. Contrarily to the deterministic case ($\sigma=0$), Fickian diffusion does not seem to induce a spatial homogenisation. In both cases, the regions with lower diffusion (i.e., where $D(x,t)$ is small) are associated with more contrasted values of the solution. Of course, one may expect differences between the covariance functions associated with these two operators, and some parametrizations may also lead to graphically visible differences, but further theoretical studies are needed to clarify these aspects. 

\end{document}